%% file: IsingOnS2.tex
\documentclass[12 pt]{article}
\newcommand\ignore[1]{}
\usepackage[left]{lineno}
\usepackage{amsmath}
\usepackage{amssymb}
\usepackage{cancel}
 \usepackage{graphicx}
 \usepackage{float}
\usepackage{braket}
\usepackage{authblk}
\usepackage{caption,subcaption}
\usepackage{comment}
\usepackage{enumitem}
\usepackage{tikz}
\usepackage{color}

\graphicspath{{fig/}}
\usepackage{amsmath}
\usepackage{amssymb}
\usepackage{graphicx}
\usepackage[margin = .75 in]{geometry}
\usepackage[pdftex, pdfstartview={FitH}, pdfnewwindow=true, colorlinks=false, pdfpagemode=UseNone]{hyperref}

\newcommand\be{\begin{equation}}
\newcommand\ee{\end{equation}}
\newcommand\bea{\begin{eqnarray}}
\newcommand\eea{\end{eqnarray}}
\newcommand{\<}{\langle}
\renewcommand{\>}{\rangle}

\newcommand\mR{ \mathbb R}

\newcommand\mS{ \mathbb S}

\usepackage{lineno}
\begin{document}

\clearpage\thispagestyle{empty}
\begin{center}

{\Large \textsc{ The Ising Model on $\mS^2$}}

\vskip15mm

Richard C. Brower$^{1}$   and Evan K. Owen$^1$

\vskip5mm


\it{$^1$Department of Physics, Boston University, Boston, MA 02215-2521, USA}\

\vskip5mm

\tt{brower@bu.edu, ekowen@bu.edu}

\end{center}


\begin{abstract}
  We define a 2-dimensional Ising model on a triangulated sphere,
  $\mS^2$, designed to approach the exact conformal field theory (CFT)
  in the continuum limit.  Surprisingly, the derivation leads to a set
  of geometric constraints that the lattice field theory must
  satisfy. Monte Carlo simulations are in agreement with the exact
  Ising CFT on $\mS^2$.  We discuss the inherent benefits of using
  non-uniform simplicial lattices and how these methods may be
  generalized for use with other quantum theories on curved manifolds.
 \end{abstract}

\setcounter{tocdepth}{2}
\setlength{\parskip}{.07in}
\tableofcontents
\setlength{\parskip}{.2in}
\newpage

\setcounter{page}{1}
\section{Introduction}
\label{sec:Intro}

Lattice Monte Carlo has proven to be a powerful method to numerically solve
 non-perturbative quantum field theories  for condensed matter
and relativistic high energy physics -- most celebrated in the QCD
sector of the standard model.  But for the most part the high precision
results ~\cite{Aoki_2022} are restricted to flat Euclidean space
discretized on hypercubic lattices.  Extensions to curved manifolds
would open up new frontiers. For example the simulation of conformal
~\cite{Brower2020RadialLQ}  or near conformal
theories~\cite{Appelquist_2010} on discretization of a
Riemann~\cite{Brower2018LatticeF} sphere $\mS^d$ or on a cylinder,
$\mR \times \mS^{d}$ for radial quantization ~\cite{Rychkov:2016iqz}.
For conformal field theories, unlike toroidal lattices,  the finite volume errors for spatial images are removed, giving
direct access to conformal parameters. 

However the problem with curved manifolds, even for  spherical manifolds in  2 or more dimensions ~\cite{ Cardy:1985xx},  is that there are  only a finite number of uniform discretizations.  For the 2
sphere there are 5 Platonic solids.  The largest, the icosahedron,
consists of 20 equilateral triangles invariant under   the finite 120
element subgroup, $A_5\times Z_2$, of $O(3)$.  For the 3 sphere, the largest uniform
sub-manifold  consists of 600 equilateral tetrahedrons in  the 14400
element Coxeter group $H_4$ subgroup of O$(4)$.  Similar restrictions
apply to Anti-de Sitter
space~\cite{PhysRevD.103.094507,PhysRevD.105.114503}. This means that
for a uniform discretization on constant curvature  manifolds,
there is a fundamental minimum lattice spacing relative to the
curvature. To avoid this limitation we seek  a method for simulating a
theory on a non-uniform lattices, which exactly approaches the symmetries of a
smooth manifold as the effective lattice spacing goes to zero.   

For classical physics, expressed as a system of partial differential
equation, the problem of discretization on curved manifolds is solved
by the Finite Element Method (FEM)~\cite{Desbrun2005DiscreteEC}.
Solutions to the discrete equations of motion on piecewise flat
manifolds, compose of d dimensional simplicies (e.g. 2d triangles, 3d
tetrahedrons, etc.), converge to the exact continuum on a suitable
sequence of refined lattices.  The same simplicial lattices were also
introduced in 1961 for the metric field in Regge Calculus
(RC)~\cite{Regge1961GeneralRW} to discretize the Einstein-Hilbert
action resulting in powerful tool to construct numerical solutions to
Einstein gravity.  However {\bf quantum fields} on these discrete
manifolds are not so forgiving. Finite element methods fail.

One way to see the difficulty for quantum fields theories is to note
that the ultraviolet divergences are sensitive to lattice cut-off,
which is  no longer  independent of position.  As a consequence  in
general there is no global second order fixed point. Removing the UV cut-off 
fails to converge to the quantum field theory or even to give the
continuum renormalized perturbation theory.  Our first remedy
implemented in the Quantum Finite Elements (QFE)
project~\cite{Brower:2016vsl} was to modify the FEM discretization by
additional quantum counter terms to cancel the local cut-off dependence of UV
divergent perturbation loops. Implemented for the super renormalizable
$\phi^4$ theory enabled accurate lattice simulations for CFT data for
2d $\phi^4$ on $\mS^2$~\cite{Brower2018LatticeF} and for 3d on
$\mR \times \mS^2$~\cite{Brower2020RadialLQ,ayyar2023operator}.
However subsequent studies suggested that exact results for lattice $\lambda_0\phi^4$ theory in the continuum limit also required tuning the bare dimensionless coupling  ($\lambda_0$) to zero by  fixing UV renormalized  parameters.

Here we seek a more general approach, applicable  to strong coupling
infrared  fixed points. To this end, we studied the
universally equivalent lattice Ising model. Surprisingly we found a
solution for the 2d Ising model on a triangulated sphere with only
nearest neighbor couplings.  By smoothing the simplicial 
geometry, combined with matching the lattice coupling, this enabled simulations
to reach the CFT in the continuum.  A first step to this matching
condition was identified in
Ref.~\cite{Brower_2023} in the analysis of the Ising model in flat
space as a function of a global affine transformation. To restore
Poincar\'e invariance of quantum correlators  fixes the  map
from the geometry (edge lengths of triangles) to the couplings on the
triangular Ising lattice as function~(\ref{eq:uniform_couplings}).  This  map
was found analytically  using the star-triangle
relation~\cite{Pokrovsky1982StartriangleRI,Baxter1982ExactlySM} and
the free fermion sector of the trivalent dual Ising model~\cite{Wolff2020IsingMA}.

In Sec.~\ref{sec:Generalize} ,  we
show how to generalize flat affine  analysis  to any  triangulation
of the  Riemann sphere.  The Kramers Wannier transformation
maps the  triangulated  Ising model  to a dual Ising model on
a trivalent lattice, which is subsequently  mapped to free Wilson-Majorana
fermions~\cite{Wolff2020IsingMA}.  All three representations give
equivalent  correlators.   With  suitable
smoothness assumptions, the fermion representation  is
used to determine the geometry at short distances  on  each tangent plane of the curved manifold.   The
reader may prefer first to skip the technical
details in subsections.~\ref{sec:KWduality} - \ref{sec:WMcontinuum},
returning when needed.  In
Sec.~\ref{sec:IsingSphere}, this map is  applied to our
target application on $\mS^2$ as a sequence of
 refined lattices approaching  the continuum.
Extensive Monte Carlo simulations
agree with  the exact  Ising
minimal $c = 1/2$  CFT  two point function on the sphere.  In Sec.~\ref{sec:Conclusion} we discuss 
further tests and generalizations.  We conjecture that a local affine
transformation to the tangent planes is sufficient for any lattice field
on a smooth manifold, when  implemented with a suitable ancillary numerical algorithm.

 \section{Geometry of  2d Ising Graphs}
\label{sec:Generalize}

The heart of the problem for lattice field theory on curved manifolds is how to  reconcile two conflicting representations of geometry.
In Regge calculus, the   geometry consists of  a collection of simplicies (e.g.  triangles in 2d) with flat
interiors. The edge lengths, $\ell_{ij}$,  on the graph  are a replacement for metric, $g_{\mu\nu}(x)$. Hence the title of 
Regge's seminal 1961 paper~\cite{Regge1961GeneralRW} on {\tt General relativity without
  coordinates}!   The continuum requires taking edge lengths to zero
and re-introducing a differentiable 
metric $g_{\mu \nu}(x)$. G. Feinberg, R. Friedberg, T. D. Lee, and
  H. C. Ren~\cite{Feinberg:1984he}
  give a detail procedure applied to spherical manifolds.  In contrast
  a  lattice quantum field theory on this {\bf same} simplicial graph  is defined by 
local  dimensionless couplings on the edges, $K_{ij}$, with no
metric. The quantum field geometry emerges  dynamically  
determined by computing correlators as one approaches a second order
phase boundary.  To reconcile Regge geometry and
lattice fields on  a curved manifold
  requires finding  a map coupling  constants, $K_{ij}$,  and  edge
  lengths, $\ell_{ij}$,
  \be
\mbox {Map:} \\ \quad K_{ij}  \rightarrow \ell_{ij} \; ,
  \ee
  consistent with the target continuum geometry.

\begin{figure}[h]
  \begin{center}
\includegraphics[width=0.5\textwidth]{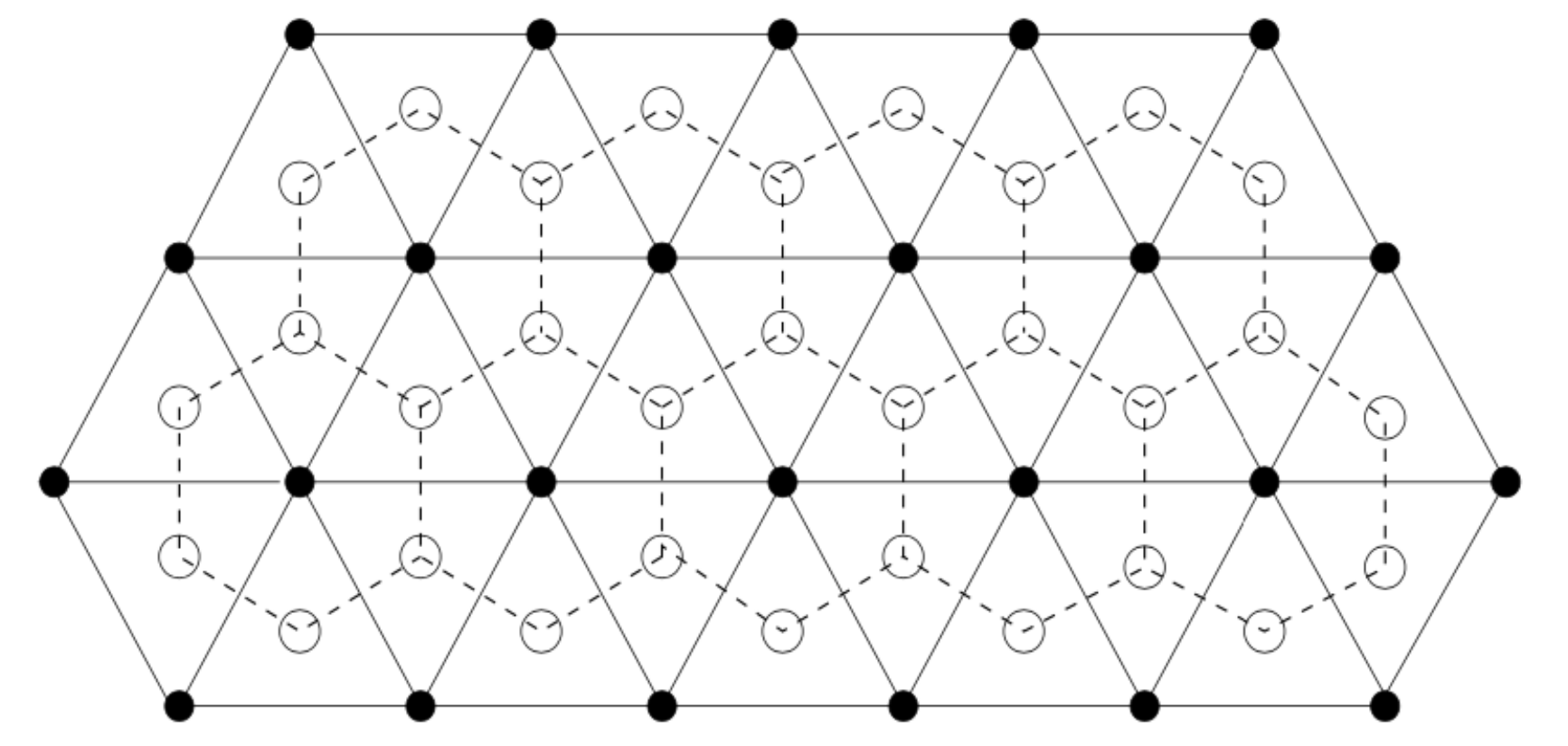}
\input{Figures/Duality2_rcb}
\caption{On the left is the triangular graph   (with black dots, solid lines) and its  hexagonal dual graph (with open circles, dashed lines). In  Regge Calculus 
the geometry is determine by assigning   edge length, $\{\ell_{ij}\}$, to the each simplex.  On the right
for affine  flat space  example, the lattice
action has 3 independent couplings, $\{K_1,K_2,K_3\}$  identical  triangles   3 edge lengths, $\{\ell_l,\ell_ 2, \ell_3\}$. 
}
\label{fig:graphsDuality}
\end{center}
\end{figure}

Interestingly this problem already arises in flat space for the 2d Ising model
studied in \cite{Brower_2023}. An affine transformation
on an equilateral triangle 
takes it to a general simplex  with 3
distinct edge lengths, $\ell_l,\ell_ 2, \ell_3$. For a scalar field in this uniform affine lattice,  it is reasonable to
introduce 3 corresponding couplings, $K_1, K_2, K_3$ as illustrated on
the right in Fig.~\ref{fig:graphsDuality}. This preserves the discrete
translational invariance of on the lattice. 

For free  scalar field theory with lattice action:
$S[\phi_j] = \sum_{\hat i= 1,2,3} K_i (\phi_{j + \hat i}  -
\phi_j)^2$, 
linear finite elements provides the
correct map,
\be
2K_i = \frac{\ell^*_i}{\ell_i} \; .
\label{eq:freeMap}
\ee
The couplings are expressed as the ratio of Voronoi dual lengths, $\ell^*_i$, between
circumcenter of two  adjacent triangles with common edge length $\ell_i$.
This restores rotational symmetry in the continuum for the
resultant FEM Laplace-Beltrami operator. Incidentally
this is  an example of
the elegant Discrete Exterior Calculus~\cite {Desbrun2005DiscreteEC}
(DEC) which applies to general  simplicial manifold
in any dimension.

 In contrast  for  the Affine Ising lattice~\cite{Brower_2023} 
  the correct map is
 \begin{equation}
\label{eq:uniform_couplings}
\sinh 2 K_i  = \dfrac{\ell_i^*}{\ell_i} \; ,
  \end{equation}
 to restore rotational symmetry for critical Ising CFT on $\mR^2$. 
This  is a non-perturbative quantum effect  at the Wilson Fisher fixed
point,  not part of  classical  FEM methods.   The map in
Eq.~(\ref{eq:uniform_couplings}) also fixes the critical surface at 
\be
p_1 p_2 + p_2 p_3 + p_3 p_1 = 1 \quad, \quad p_i =e^{ - 2 K_i} \; .
\label{eq:CriticalSurface}
\ee
Both are consequence of the star-triangle
identity~\cite{ Pokrovsky1982StartriangleRI,Baxter1982ExactlySM}
between the triangular lattice with couplings $K_1, K_2, K_3$  and its Kramers Wannier dual with
with couplings $L_1,L_2,L_3$ fixed by $e^{-2 K_i} = \tanh(L_i)$. The geometry is then fixed by the loop 
expansion for free Wilson-Majorana  fermions on the trivalent dual lattice.  

The take away is that even on $\mR^2$, restoring Poincar\'e
invariance require solving the the quantum theory.  The problem we now face is how to generalize this
to curved manifolds, forgoing  the dependance on analytical tools of the  2d flat
space Ising model. The hope is that the map in Eq.~(\ref{eq:uniform_couplings}) can  still be apply locally on each tangent plane as we take the continuum limit.

\subsection{Kramers Wannier Duality}
\label{sec:KWduality}

A general triangulated surface  with no boundaries is defined by a set of $N$ vertices, $E$ edges, and $F$ triangular faces  satisfying  Euler's theorem, $F - E + N = 2 -2 g$. The topology is fixed by the genus , $g$.
Constant curvature metric exist for the positive curvature
sphere at $g=0$, zero curvature torus at $g = 1$ and
negative curvature hyperbolic Riemann surfaces  for  $g > 1$.  

\begin{figure}[h]
    \centering
    \includegraphics[width=0.6\textwidth]{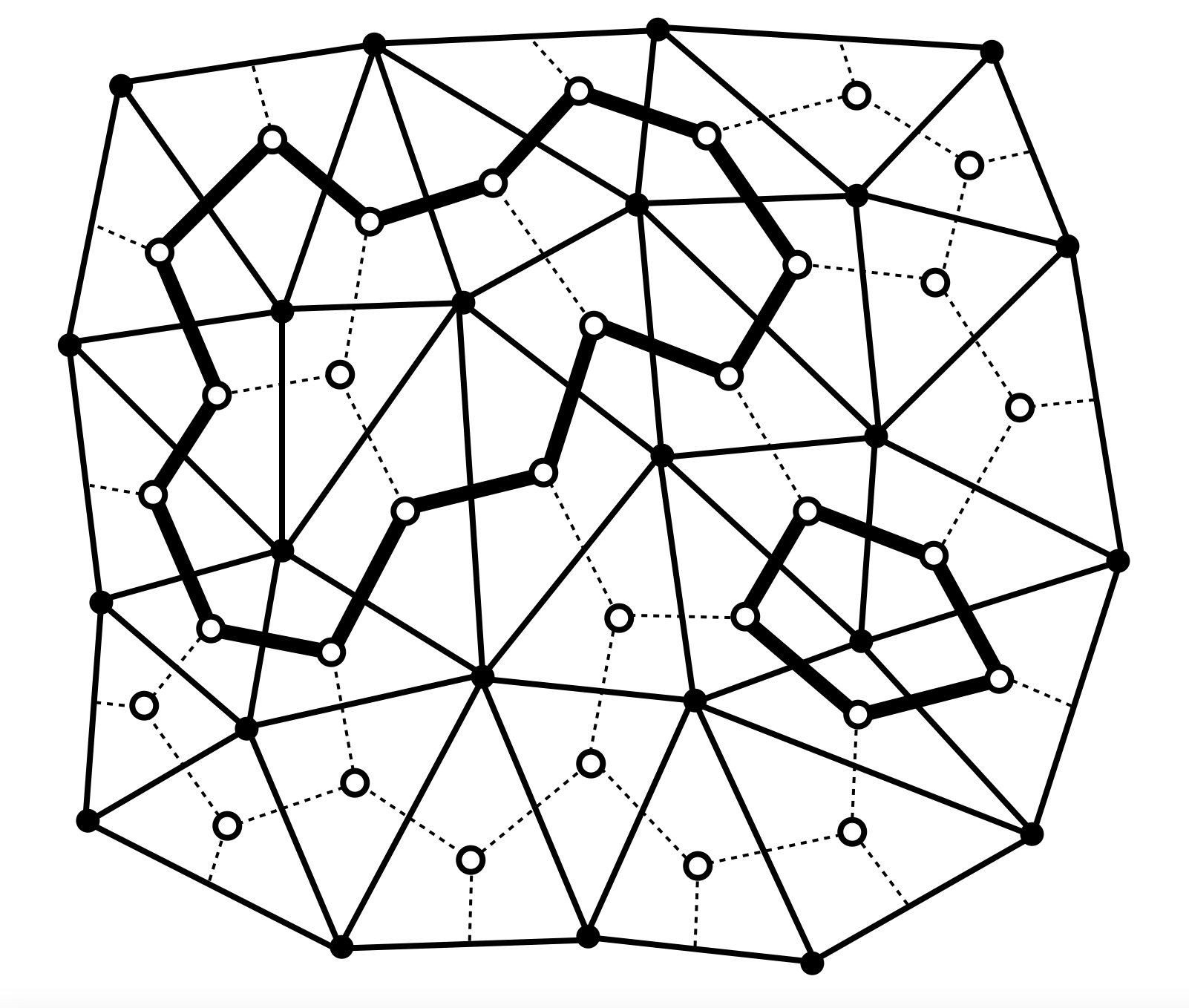}    \caption{An
      illustration of a loop configuration for a non-uniform
      simplicial lattice and its trivalent dual lattice on portion of
      Riemann manifold. The bold solid lines show two  dual loops ($\Gamma_i$)   in the low temperature.   }
    \label{fig:ising_loop}
  \end{figure}

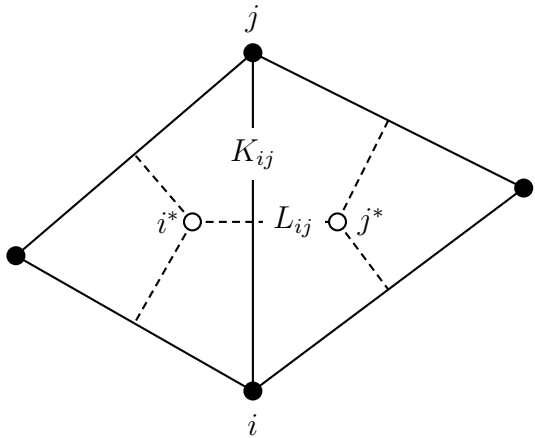
\begin{figure}[h]
    \centering
    \input{Figures/SimplicialCouplings}
    \caption{Couplings for a single edge on both the simplicial lattice (closed circles, solid lines) and its trivalent dual (open circles, dashed lines).}
    \label{fig:simplicial_couplings}
  \end{figure}

The graph dual to any triangulated Riemann surface is  trivalent  graph as illustrated in Fig.~\ref{fig:ising_loop}
with a one to one match of edges and dual edges.
The Ising model triangular graph can be define the sum over the 
edges of the triangulated graphs with action, 
\begin{equation}
\label{eq:tri_action}
    S_{\triangle} = - \sum_{\langle ij \rangle} K_{ij} s_i s_j \;.
  \end{equation}
  Similarly on the  dual graph we
  can define a dual   Ising spin system 
  \begin{equation}
  \label{eq:dual_action}
    S_{\text{dual}} = - \sum_{\langle ij \rangle} L_{ij} s_{i}s _{j}  \; .
  \end{equation}
  Properly the dual Ising spins are now disorder  variables,
  $\mu_{i^*}= \pm 1$, at dual sites $i^*$ but for notational simplicity
  we   label them  by $\mu_{i^*} 
  \rightarrow s_{i} $  without the star on dual sites  $i^*
  \rightarrow i$.  The coupling  constants for the triangle  and
  trivalent dual Ising models are  paired as illustrated in Fig.~\ref{fig:simplicial_couplings}. In passing  we
  note the special simplicity  the trivalent
  dual of 2d triangle simplicial complex for any  Riemann triangulated surface is
  at the heart of the random graph solution to 2d string theory as
   a large N matrix model in the double scaling limit~\cite{FRANCESCO19951}.
  
For any fixed  triangulation,  there is a general Kramers Wannier map~\cite{PhysRev.60.252}  found by comparing strong and weak loop expansions. 
On the triangular graph, the   {\bf low temperature}  expansion  is  a power series in the number of broken bonds,
\be
\label{eq:tri_kramers}
Z_{\triangle}[K] = Tr[e^{K_{ij} s_i s_j} ] = \sum_{\{s_i\}= \pm 1} \prod_{\langle ij \rangle} e^{K_{ij}}  [ e^{ -2  K_{ij}} + (1 - e^{ -2  K_{ij}} )\delta_{s_i, s_j} ] 
=  2 \prod_{\langle ij \rangle}  e^{K_{ij} } \sum_{\{\Gamma\}} \prod_{\langle ij \rangle \in \Gamma} e^{-2 K_{ij}}
\ee
 Since the  dual lattice is trivalent the path of broken bounds form non-intersecting loops and $\Gamma$ enumerates the
 loops ensemble. 
 
These same loop enumeration also give the   {\bf  high temperature}
for  aligned spins $s_i = s_j$ on  the dual lattice for  sphere ($g =0$) or the topologically equivalent $CP(1)$ complex plane,
\be
\label{eq:dual_kramers}
Z_{\text{dual}}[L]  = Tr[e^{L_{ij} s_i s_j} ]  =  \sum_{\{s_i\}=\pm 1} \prod_{\langle ij \rangle} \cosh L_{ij}(1 + s_i s_j \tanh L_{ij}) 
=   2^F \prod_{\langle ij \rangle} \cosh L_{ij}   \sum_{\{\Gamma\}} \prod_{\langle ij \rangle \in \Gamma} \tanh L_{ij}
\ee
where $F$ is the number of faces on the triangular sphere.  For $g =0$,   partition functions in
Eqs,~(\ref{eq:tri_kramers}-\ref{eq:dual_kramers}) are equivalent if
we assert the identity,
\begin{equation}
  e^{-2 K_{ij}} = \tanh L_{ij}
  \label{eq:dualMap}
\end{equation}
or its equivalent to the symmetric form
\begin{equation}
\label{eq:dual_couplings}
    \sinh 2 K_{ij} \sinh 2 L_{ij} = 1 \; .
  \end{equation}
  Although we use the term "expansion", both Eq.~(\ref{eq:tri_kramers}) and Eq.~(\ref{eq:dual_kramers}) are exact. 
  We note that for higher genus Riemann surfaces ($g \ge 1$) such as the modular torus, the non-contractible loops require addition care with boundary conditions.

This is a powerful map central to  our construction. It fixes
  the ratio of the partition functions  or  equivalently the difference of the free energies ,
\be
F_\triangle[K]  =   F_{\text{dual}}[L]   + \frac{1}{2} \sum_{\<i,j\>} \log(\sinh(2 L_{ij})) + (F - 1)\log(2)  \; .
\ee
Taking derivatives with respect to the link coupling,  $K_{ij}$,   
all correlation function for  $Z_2$  even  Ising  energy   operators on the edges of  triangular lattice are  mapped to  dual correlation
functions. For example if we consider the connected two point energy-energy correlators between any two edges $\<1,2\>$
and  $\<3,4\>$, we have
\be
\partial_{K_{12}} \partial_{K_{34}} F_\triangle[K]= \< (s_1 s_2) (s_3 s_4)\>_\triangle
\ee
which is mapped to the dual by the identity, 
\be
\< (s_1 s_2) (s_3 s_4)\>_{\text{dual}}  = 
\sinh(2 L_{12}) \sinh(2 L_{34}) \;  \< (s_1 s_2) (s_3 s_4)\>_\triangle 
\label{eq:KLidentity}
\ee
using the chain rule, it $d L/d K  =  - \sinh(2 L)$ on Eq.~(\ref{eq:dual_couplings}).

At present we don't have enough information to determine the geometry of the
quantum system.  The positions of the vertices, the lengths of the
edges, and the locations of the dual sites within the triangular faces
are all unknown quantities. Only if the lattice model has a well-defined
continuum limit at a second order phase boundary, can
we find the emergent quantum geometry
through  its correlation functions. To implement
this,  we now map the trivalent Ising partition function  to  an equivalent 
 Wilson-Majorana free fermion partition function.

\subsection{Wilson-Majorana fermion loop expansion}
\label{sec:WilsonMajorana}

\begin{figure}
    \centering
    \includegraphics[width=0.5\textwidth]{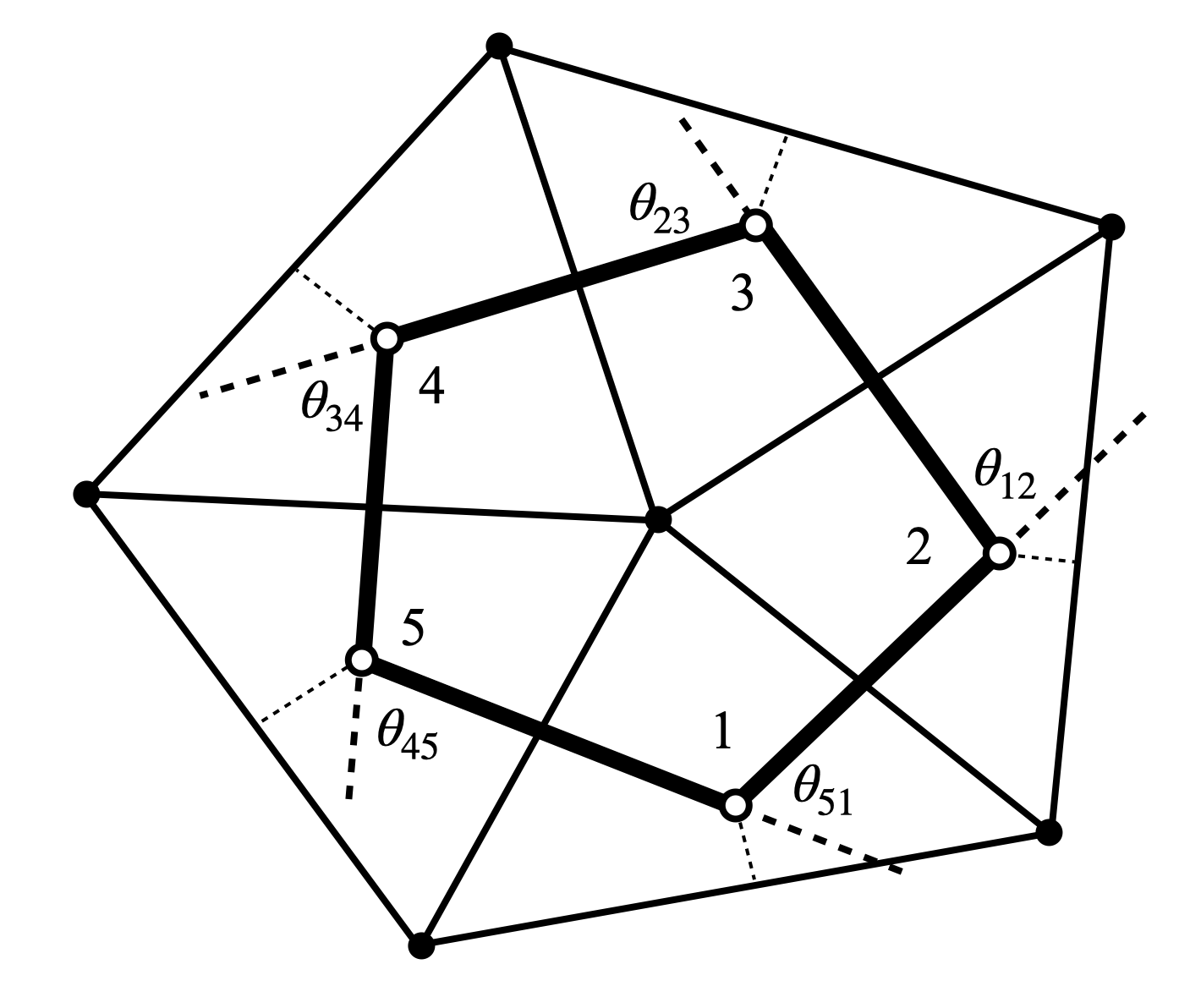}
    \caption{The  angles at each of the trilinear
    sites provide a  discrete rotation of the Wilson-Majorana propagator. This forms a Voronoi  polyhedron dual  to the central vertex ($\bullet$). with area  $A^*_ \bullet$. }
    \label{fig:phase_angles}
  \end{figure}

We introduce Wilson-Majorana fermion  on the trivalent dual lattice,
with fields $\psi_i^{\alpha}$  ($\alpha=1,2$). Each
field is a  two-component spinor that obeys the
charge conjugation constraint
$\bar \psi_i^{\beta} = \psi^{\alpha}_i \epsilon^{\alpha \beta}$,
i.e. $\bar \psi^1_i = -\psi^2_i$ and $\bar \psi^2_i = \psi^1_i$.
The action is 
\begin{equation}
\label{eq:fermion_action}
    S_{\psi} = \dfrac{1}{2} \sum_i \bar \psi_i \psi_i - \sum_{\langle ij \rangle} \kappa_{ij} \bar \psi_i P_{ij} \psi_j \; ,
\end{equation}
with local  hopping parameters, $\kappa_{ij}$ and 
Wilson spinor propagator,  
\begin{equation}
    P_{ij} = \dfrac{1}{2} (1 + \hat e_{ij} \cdot \vec \sigma) \; .
\end{equation}
 The vectors,  $\hat e_{ij}$, are defined as a unit vector on the dual links from $i$ to $j$, so that Wilson factor are projection operator, $ P^2_{ij} = P_{ij}$, with two eigenvalues $0,1$. Wilson-Majorana fermion with real Grassmann  variables  has a symmetry property
identifying directed links $ i\rightarrow j \equiv \{ i,j \}$ with the reflected direction $ j\rightarrow i \equiv \{ j,i \}$ :
 \be
  \bar \psi_i( 1 + \hat e_{ij} \cdot \vec \sigma) \psi_j =
  \bar \psi_j( 1 + \hat e_{ji} \cdot \vec \sigma) \psi_i
 \label{eq:Reflection}
  \ee
  where $\hat e_{ji} = - \hat e_{ij}$.   This is an additional reflection symmetry,  independent of so called $\gamma_5$ hermiticity for Wilson Fermions. So there is no need to add links in both directions.   This is similar to the
  convention introduced in Ref.~\cite{Brower:2016vsl} for Wilson
  fermions,  except now the  fermions are moved from the  triangulated simplex to sites on
  the   dual lattice.  The geodesics between dual sites are  
straight lines illustrated in Fig.\ref{fig:phase_angles}.

 Now we show that   free fermion partition function,
\begin{equation}
\label{eq:fermion_partition_fn}
Z_{\psi} = \int \mathcal{D} \psi e^{-S_{\psi}} \; ,
\end{equation}
with  Grassmann path integral measure
\begin{equation}
    \mathcal{D} \psi \equiv \prod_i d \psi_i^1 d \psi_i^2
  \end{equation}
   with  appropriate hopping parameters: $\kappa_{ij}$  is equivalent to the Ising partition
  function on the dual lattice (\ref{eq:dual_kramers}). We can expand
  the exponential in Eq.~(\ref{eq:fermion_partition_fn}),
\begin{equation}
    Z_{\psi} = \int \mathcal{D} \psi_i \prod_i  \left( 1 -
      \dfrac{1}{2} \bar \psi_i \psi_i \right) \prod_{\langle ij
      \rangle} \left( 1 + \dfrac{1}{2} \bar \psi_i (1 + \hat e_{ij}
      \cdot \vec \sigma)  \psi_j \right) \; ,
  \end{equation}
 discarding all terms above first order which vanish in a Grassmann
 integral. Now, following the standard rules for integration of Grassmann variables, the only nonzero terms are those which include every pair $\bar \psi_i \psi_i = \epsilon^{\alpha\beta} \psi^\alpha_i\psi^\beta_i$ exactly once
on each site $i$. The set of nonzero terms is exactly described by the
same set of unique loop configurations $\{\Gamma\}$ that we used in
Sec. \ref{sec:KWduality}. Due to Eq.~(\ref{eq:Reflection}),  it does
not matter which direction we choose to traverse each loop, the result
will be the same either way. Discarding irrelevant constant factors, the fermion partition function becomes
\begin{equation}
    Z_{\psi} = \sum_{\{\Gamma\}} (-1)^n \operatorname{tr} \left[ \prod_{\langle ij \rangle \in \Gamma} \dfrac{1}{2} \kappa_{ij} (1 + \hat e_{ij} \cdot \vec \sigma)  \right]
\end{equation}
where $n$ is the number of closed loops in $\Gamma$.

The evaluation  depends  on  the product of adjacent projection operators.  We may choose an arbitrary coordinate system with $\hat e = [e_x , e_y]$  across
each edge, given as a  rotation from the dyadic along the x axis
\be
P(\vec n) = \frac{1}{2} (1 + \hat n \cdot \vec \sigma)  = e^{\textstyle i \theta \sigma^z/2} \big[ \frac{1  + \sigma^x}{2}  \big] e^{\textstyle- i \theta \sigma^z/2}  
\ee
or  introducing complex variable $z = n_x + i \; n_y =e^{i \theta}$  as  dyadic, 
\be
  P(\theta)  = \frac{1}{2}
   \begin{bmatrix}
  e^{ i\theta/2} & \\
   0 &  e^{ - i\theta/2}
\end {bmatrix}
  \begin{bmatrix}
  1 & 1\\
  1 & 1
\end {bmatrix}
  \begin{bmatrix}
  e^{- i\theta/2} & \\
   0 &  e^{  i\theta/2}
\end {bmatrix}
=  \frac{1}{2}
\begin{bmatrix}
 \sqrt{ z }\\
  \sqrt{z^*}  
\end{bmatrix}
\begin{bmatrix}
  \sqrt{z^*}  & \sqrt{z}
\end{bmatrix} \equiv |z \> \< z| \; ,
\ee
where have used  the  constraint $z z^* = 1$. In the loop expansion each
corner has the spinor matrix, $\< z_i |z_j\> =  e^{ \theta_{ij} \sigma^z/2}$ which is 
evaluated between the non-zero  Grassmann integral  $\epsilon^{\alpha \beta} \psi^\alpha \psi^\beta$. This takes the trace,  $Tr[ e^{ \theta_{ij} \sigma^z}/2] = 2 \cos(\theta_{ij}/2)$, resulting in the loop expansion,
\begin{equation}
\label{eq:fermion_loop}
    Z_{\psi} = \sum_{\{\Gamma\}} \prod_{\langle ij \rangle \in \Gamma} \kappa_{ij} \cos \dfrac{\theta_{ij}}{2} \;.
  \end{equation}

Here our notation is a bit compressed.  Fermions live on dual sites $i$ and propagate on directed links, $i \rightarrow j \equiv \{i,j\}$, along unit vectors $\hat e_{ij}$ with couplings $\kappa_{ij}$.  For a fixed  loop as illustrated in  Fig.~\ref{fig:phase_angles} we may  relabel the sites in sequence a $1\rightarrow 2 \rightarrow 3 \rightarrow \cdots$  with angles $\theta_{i+1,i}$
  at each vertex $i$.  Of  course the vertex angles depend  on the path. 
  So a proper notation
  {\bf without relabeling sites}  for the rotation at 
   vertex  $k$  for a general path $i\rightarrow k \rightarrow j $ is determined by the scalar product
  $\hat e_{ik}  \cdot \hat e_{kj} = \cos(\theta_{ik} -\theta_{kj})$.
  This requires
  3 site labels for the rotation angle:  $\theta^k_{ij} \equiv \theta_{ik} -\theta_{kj}$. The sign is undetermined 
  because the loop is unchanged if we reverse the direction due the reflection identity in Eq.~(\ref{eq:Reflection}).

\begin{figure}[h]
    \centering
    \input{Figures/GenericIsing}
    \caption{Definition of the 6 angles used in Eq.~(\ref{eq:fermion_ising_match}).
     The open circles are the circumcenters of the two triangles.}
    \label{fig:generic_ising_angles}
  \end{figure}
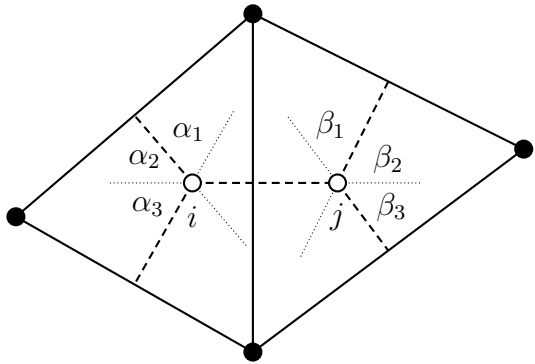

    In the affine case  there are 3 distinct  links enumerated by  3 out-going (or in-going) vectors, 
     $\hat n_1, \hat n_2, \hat n_3$  in a bipartite graph.  In Ref.~\cite{Brower_2023}, we  proved the identity  
  \be
  t_2/\kappa_2 = \frac{\cos(\theta_{12}/2)\cos(\theta_{23}/2)}{\cos(\theta_{31}/2)} = 
  \frac{\cos((\theta_1 - \theta_2)/2)\cos((\theta_2 - \theta_3)/2)}{\cos((\theta_3 - \theta_1)/2)} = \frac{\<z_2|z_1\> \<z_3|z_1\>}{\< z_2|z_3\>}
  \label{eq:affineID}
  \ee
  and cyclic permutations. Remarkably  this can be  generalized for any  trivalent lattice to the identity 
\begin{equation}
\label{eq:fermion_ising_match}
    \tanh^2 L_{ij} = \kappa_{ij}^2 \dfrac{\cos (\alpha_{2}/2) \cos (\alpha_{3}/2) \cos (\beta_{2}/2) \cos (\beta_{3}/2)}{\cos (\alpha_{1}/2) \cos (\beta_{1}/2)}
\end{equation}
where the six angles are defined in Fig. \ref{fig:generic_ising_angles}. 
It follows by matching terms in loop expansion (\ref{eq:dual_kramers}) and (\ref{eq:fermion_loop}) or  proven algebraically by extending the  analysis above for each of the 4 paths through the link $\<i,j\>$  with $\alpha_2$ or  $\alpha_3$ on the left and $\beta_2$ or  $\beta_3$ on the right.  In flat space this  simplifies 
to square of the affine identity in Eq.~(\ref{eq:affineID}),  with the
condition  that  $\alpha_i = \beta_i$. We can use
Eq.~(\ref{eq:dual_couplings}) to relate the triangular lattice coupling $K_{ij}$ to the corresponding dual lattice coupling $L_{ij}$ and by this association we now have three different models with equivalent partition functions.

\subsection{Lattice continuum limit}
\label{sec:WMcontinuum}

We have now shown that the Ising model on  the simplicial triangular lattice, its trivalent dual and the free Wilson-Majorana fermion all have equivalent partition functions, and therefore they all describe systems with equivalent dynamics. But we haven't yet identified the critical point of the system, nor have we determined the appropriate geometry of the continuum theory (assuming such a theory exists). This requires
a smooth local lattice theory, most likely close to the uniform flat space triangulated lattice
studied in Ref.~\cite{Brower_2023}.

The first step is we need to have a smooth definition of the scalar curvature  that
at present are delta function singularities at the triangular manifold.  This can be
done to  $O(a^2)$ using the theorem in the remarkable paper
on {\tt Lattice Gravity Near the Continuum Limit} ~\cite {Feinberg:1984he} that proves that in 2d the scalar curvature, $R =2  \epsilon_{i}/A^*_i + O(a^2)$,  is given by
the ratio of the deficit angle $\epsilon_{i^*} $ at the vertices of
the triangulated manifold and the dual area $A^*_i$
on the trivalent lattice as illustrated in Fig. \ref{fig:phase_angles}. 
Thus we can we can introduce a smooth approximation
to the manifold 
by  spreading  the singular curvature uniformly over the dual area and 
introducing a modified fermion action analogous to the 
Dirac fermions in Ref.~\cite{Brower:2016vsl}  for the simplicial lattices, 
\begin{equation}
\label{eq:smooth_fermion_action}
   \widetilde S_{\psi} = \dfrac{1}{2} \sum_i \bar \psi_i \psi_i -\dfrac{1}{2} \sum_{\langle ij \rangle} \kappa_{ij} \bar \psi_i ( 1 + \hat e_{i j} \cdot\vec \sigma )\Omega_{ij}\psi_j \; .
  \end{equation}
  Now $\hat e_{i j}$ is a unit tangent vector at site $i$ on the geodesic from $i$ to $j$ and $\Omega_{ij} = \Omega^\dag_{ji}$
  is the rotation (or discrete spin connection) to the vector $\hat e_{ji} $ at site $j$ on the reverse geodesic to $i$.
  Due to local curvature there are two distinct
  tangent vectors at each end of each link  that are no longer  reflected: $\hat e_{ji} \ne - \hat e_{ij} $. Instead  they  obey the discrete tetra hypothesis introduced
  in  Ref.~\cite{Brower:2016vsl},
  \be
\hat e_{ji} =-  \Omega^\dag_{ij} \hat e_{i j} \Omega_{ij}  \; .
\ee
  Here the local curvature between dual lattice points   requires replacing  the
  in Eq~\ref{eq:Reflection}  by 
  \be
  \bar \psi_i ( 1 + \hat e_{i j} \cdot\vec \sigma )\Omega_{ij}\psi_j = \bar \psi_j \Omega_{ji}( 1 + \hat e_{ji} \cdot\vec \sigma )\psi_j 
  \ee
  
Now in a smooth gauge with $ \Omega_{ij} = 1 + O(a)$, we can expand (\ref{eq:smooth_fermion_action}) in the effective lattice spacing $a$ using
\begin{equation}
    \psi_j = \psi_i + \vec l_{ij}^* \cdot \vec \nabla \psi_i + \mathcal{O}(a^2)
\end{equation}
and
\begin{equation}
    \Omega_{ij} = e^{- i \vec l_{ij}^* \cdot \vec \omega} = 1 - i \vec l_{ij}^* \cdot \vec \omega + \mathcal{O}(a^2)
\end{equation}
where $\vec l_{ij}^*$ is a vector from $i$ to $j$ with length $l_{ij}^* = \mathcal{O}(a)$ and $\vec \omega$ is the continuum spin connection. The action becomes
\begin{equation}
\label{eq:fermion_action_expand}
\begin{split}
    \widetilde  S_{\psi} &= \dfrac{1}{2} \sum_i \bar \psi_i \psi_i - \dfrac{1}{2} \sum_{\langle ij \rangle} \kappa_{ij} \bar \psi_i (1 +\hat e_{ij} \cdot \vec \sigma) (1 - i \vec l_{ij}^* \cdot \vec \omega) (1 + \vec l_{ij}^* \cdot \vec \nabla ) \psi_i + \mathcal{O}(a^2) \\
    &= \dfrac{1}{2} \sum_i \left( 1 - \dfrac{1}{2} \sum_{j \in \langle ij \rangle} \kappa_{ij} \right) \bar \psi_i \psi_i + \dfrac{1}{2} \sum_{\langle ij \rangle} \kappa_{ij} \bar \psi_i (\hat e_{ij} \cdot \vec \sigma) (\vec l_{ij}^* \cdot \vec D) \psi_i
\end{split}
\end{equation}
where $\vec D = \vec \nabla - i \vec \omega$ is the covariant derivative for the spinor field and in the second line we have used $\bar \psi_i \vec \sigma \psi_i = 0$ and discarded all terms of $\mathcal{O}(a^2)$.

We would like to show that this is equivalent to the continuum action for a free fermion on a Riemannian manifold,
\begin{equation}
\label{eq:fermion_action_cont}
    S_{\text{cont}} = \dfrac{1}{2} \int d^2 x~ \sqrt{g} \bar \psi(\vec x) (m + \vec \sigma \cdot \vec D) \psi (\vec x) \;.
\end{equation}
In order to get Eq.~(\ref{eq:fermion_action_expand}) into this form, we need the $\vec \sigma$ and $\vec D$ in the lattice action to be contracted, which is not possible for an arbitrary choice of lattice vectors in a non-uniform lattice. However, provided that all adjacent triangles have the same circumradius, the following identity is true for a lattice which is the circumcenter dual of a non-uniform simplicial lattice in 2 dimensions:
\begin{equation}
    \sum_{j \in \langle ij \rangle} l_{ij}^{\mu} l_{ij}^{*\nu} = 2 A_i \epsilon^{\mu \nu}
\end{equation}
where $\vec l_{ij}$ is the lattice vector in the simplicial lattice dual to $\vec l^*_{ij}$ and $A_i$ is the area of the triangular face dual to site $i$.

In order to apply this identity, it is necessary for $\kappa_{ij}$ to be proportional to the corresponding triangular lattice length $l_{ij}$.  Substituting this all into the action we obtain
\begin{equation}
  \widetilde    S_{\psi} = \dfrac{1}{2} \sum_i A_i \bar \psi_i \left(m_i + \vec \sigma \cdot \vec D \right) \psi_i + \mathcal{O}(a^2)
\end{equation}
with a dimensionless mass parameter
\begin{equation}
\label{eq:crit_fermion_mass}
    m_i = \dfrac{1}{A_i} \left( 1 - \dfrac{1}{2} \sum_{j \in \langle ij \rangle} \kappa_{ij} \right) \;.
\end{equation}
In the limit $a \to 0$ we identify the limit of the discrete integration measure $A_i \to d^2 x \sqrt{g}$ and find that the lattice action converges to the continuum action in Eq.~ (\ref{eq:fermion_action_cont}), as required.   
To restore local rotational symmetry for the continuum 
fermion implies that the critical hopping parameter values for each link are defined by
\begin{equation}
\label{eq:kappa_crit}
    \kappa_{ij} = \dfrac{2 l_{ij}}{\sum_{k \in \langle ik \rangle} l_{ik}} \; ,
\end{equation}
    Now, in order to have a well-defined critical point, we must satisfy this relation for every link on the lattice simultaneously, but upon inspection we find that this is only possible if all of the triangular faces have equal perimeter. Combining Eq.~ (\ref{eq:fermion_ising_match}) and (\ref{eq:kappa_crit}) we obtain an expression which relates the dual lattice Ising coupling $L_{ij}$ to the geometry of the lattice
\begin{equation}
\label{eq:crit_couplings}
    \tanh^2 L_{ij} = \dfrac{4 l_{ij}^2 \cos (\alpha_{2}/2) \cos (\alpha_{3}/2) \cos (\beta_{2}/2) \cos (\beta_{3}/2)}{P_{\triangle}^2 \cos (\alpha_{1}/2) \cos (\beta_{1}/2)}
\end{equation}
where $P_{\triangle}$ is the triangle perimeter. We can also recover the critical couplings $K_{ij}$ for the triangular lattice through Eq.~( \ref{eq:dual_couplings}).

The two geometrical constraints of equal triangle circumradius and equal triangle perimeter place strict limitations on the set of simplicial lattices for which it is possible to define a critical theory with a well-defined continuum limit.  They are natural  generalizations of the flat space  affine 
solution ~\cite{Brower_2023}, which gave 
 exact Ising correlator  on flat space and  on the  genus $g =1$ modular torus.   The zero mass condition  $m_i =0$  in Eq.~(\ref{eq:crit_fermion_mass}) reduces to Eq.~ (\ref{eq:CriticalSurface}).

\section{Ising model on a 2-sphere}
\label{sec:IsingSphere}

We now proceed to test whether  these local constraints  are sufficient in the continuum to give the
critical Ising model on the Riemann sphere.

\begin{figure}[h]
    \centering
    \includegraphics[width=0.3\textwidth]{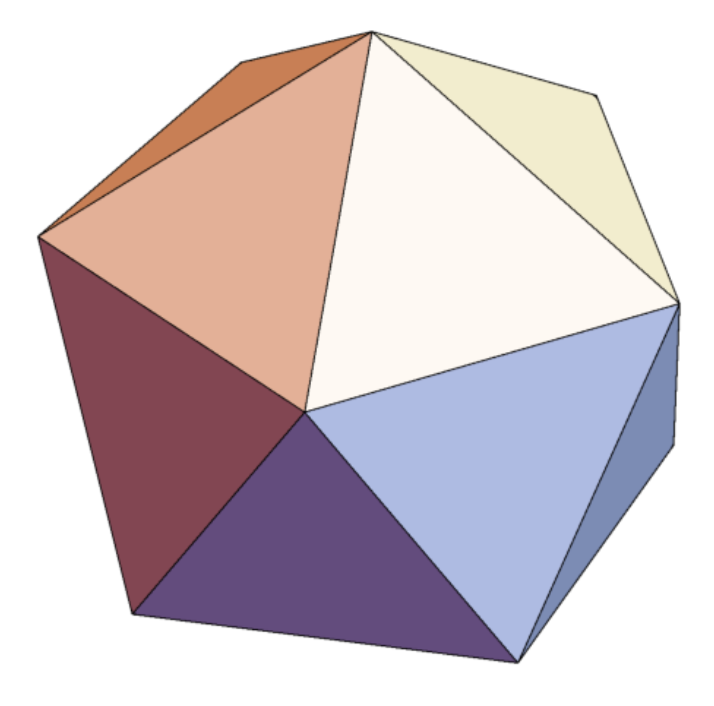}
    \includegraphics[width=0.3\textwidth]{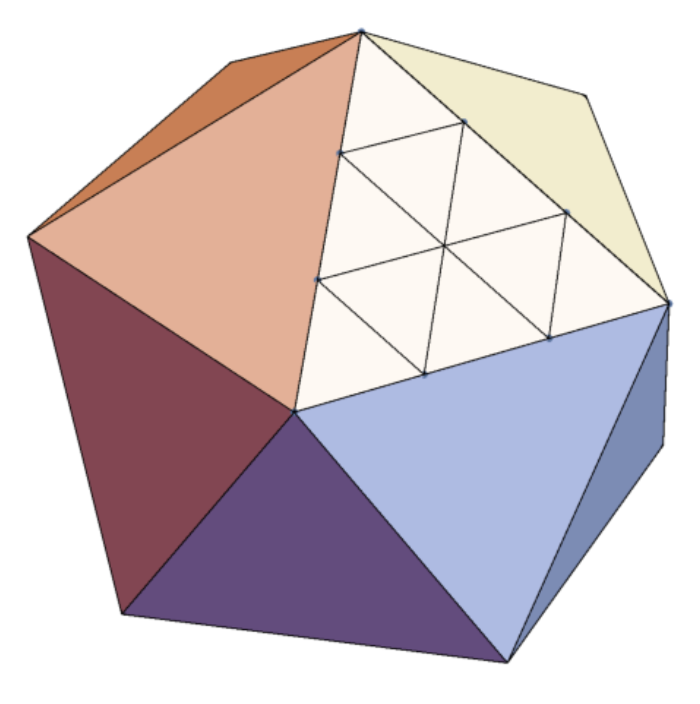}
    \includegraphics[width=0.3\textwidth]{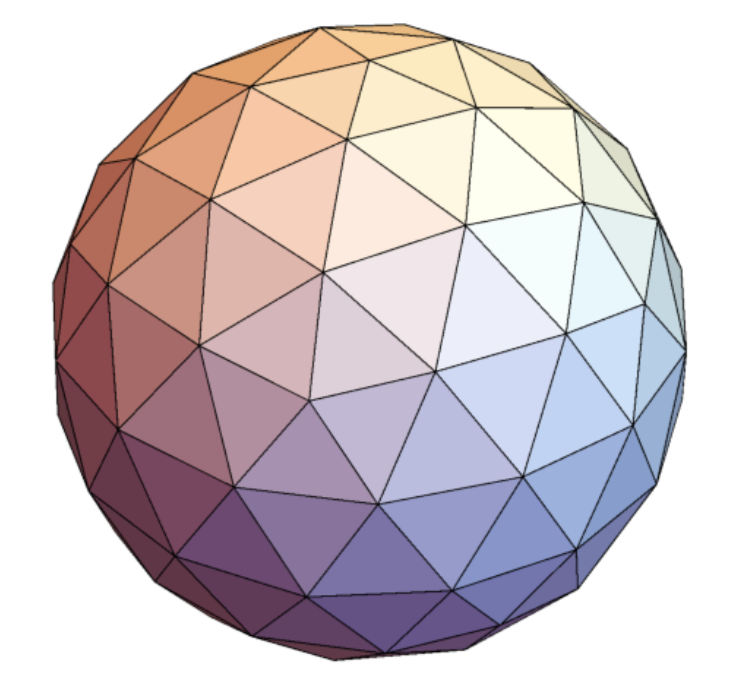}
    \caption{Steps in the basic discretization of $\mS^2$ using an icosahedral base, shown here for a refinement of L = 3 into 
    $L^2$ triangle on each of the 20 icosahedral faces subsequently projected onto the sphere. }
    \label{fig:ico_refine}
\end{figure}

We define a base discretization of a 2-sphere illustrated in
Fig. \ref{fig:ico_refine}  by
introducing equilateral triangles on each of the 20 faces of one of
the simplicial Platonic solids. These are then projecting radially
onto a unit sphere define by 3 vectors $\vec r_i$ in $\mR^3$. The result is
smooth but non-uniform triangulation of the sphere that in the
continuum limit approaches an affine map to each tangent plane.  At
finite refinement the triangulated Regge
manifold~\cite{Regge1961GeneralRW}
consist of piecewise flat triangle, with all curvature is confined
to singularities at the vertices. 

First we note that the projected  triangulated sphere
illustrate on right in Fig.\ref{fig:ico_refine}
does give critical Ising model in the continuum,
but it fails to  recover spherical isometries
for the 2d Ising CFT  on $\mS^2$.   To fix this we proceed to  move the position of the vertices on the sphere to to minimize non-uniformities in the
 triangle circumradius and perimeter as suggested in Sec.
 ~\ref{sec:WMcontinuum}. We show that  Monte Carlo simulations extrapolated to
the continuum  on these modified smooth spherical  lattices  are in good agreement with the exact
solution of the 2d Ising CFT on $S^2$. The  geometric
constraints of uniform circumradius and perimeter are indeed required
to reach the desired continuum limit quantum field theory.

We use the simplicial Ising action (\ref{eq:tri_action}) with coupling
constants,
\be
\sinh(2 K_{ij}) = \frac{\ell^*_{ij}}{\ell_{ij}} 
\ee
as a function of the edge lengths, $\ell_{ij}= |\vec r_i - \vec r_j|$,
and its circumcenter dual lengths, $\ell^*_{ij}$, perpendicular to the
edge $\<i,j\>$.  It is important to note that the dual
lattice edge lengths $l_{ij}^*$ are \emph{not} calculated by using the
geodesic distance between triangle circumcenters in the 3-dimensional
embedding space.  Under the conventions of Regge calculus, the manifold
is  defined on the flat triangular faces of the simplicial complex
so the dual lattice lengths are computed by following intrinsic Regge
geodesics as straight lines between flat triangular faces.  The only
curvature consists of delta function singularities    at the vertex
proportional to the 
deficit angle from adjacent triangles. For piecewise Regge manifolds in high
dimensions singular curvature are on 2d  co-dimension {\tt hinges}
(2d points , 3d edges, 4d triangles etc.).  Because the perimeters of pairs of triangles which
share an edge are not always equal we instead define $P_\triangle$
in Eq.~ (\ref{eq:crit_couplings}) as the geometric mean of the
perimeters of the two triangles

\subsection{Geometric uniformity}
\label{sec:Uniformity}

The solution requires two steps. One is to establish a more uniform Regge
geometric manifold to match the continuum spherical manifold.  This is
necessary to related the discrete simplicial geometry to a
differential co-ordinate on the sphere as described
Ref.~\cite{Feinberg:1984he} and utilized to go local continuum on tangent planes described in Sec.~\ref{sec:WMcontinuum} above.

The only simplicial discretizations of a 2-sphere which can be constructed from uniform triangles are the platonic solids: tetrahedron, octahedron, and icosahedron\footnote{The other two platonic solids, the cube and dodecahedron, are not included here because they are not simplicial complexes. They are, in fact, the trivalent dual graphs of the octahedron and icosahedron, respectively, while the tetrahedron is self-dual.}. We are therefore unable to construct arbitrarily refined discretizations of a sphere which satisfy the constraints of uniform circumradius and perimeter exactly. Instead, we will generate a sequence of lattices such that the non-uniformity in these two quantities goes to zero in the continuum limit.

We define the non-uniformity in the circumradius $R$ and perimeter $P$ as
\begin{equation}
    E_{R} = \dfrac{\langle R^2 \rangle}{\langle R \rangle^2} - 1 \qquad \text{and} \qquad E_{P} = \dfrac{\langle P^2 \rangle}{\langle P \rangle^2} - 1
\end{equation}
where  here the angle brackets denote an average over all triangles in the simplicial complex. These quantities can be understood as the normalized variance in the circumradius and perimeter over the entire lattice. Our goal is to move the vertices of the basic discretization (without adding or removing any edges) so that both of these quantities are minimized while retaining the point group symmetry of the original lattice. We do this by minimizing the sum $E = E_R + E_P$. In general, there are not enough degrees of freedom to find a solution such that $E=0$, however we will show that the minimum value of $E$ goes to zero as the square of the effective lattice spacing. As shown in Sec. \ref{sec:WMcontinuum}, this is sufficient to restore local rotational symmetry in the
continuum  close to each tangent plane of the critical Ising model.

\subsection{Projection for  Spherical Symmetry}
\label{sec:projSphere}

The spin-spin correlation function on the sphere is calculated by
projecting onto spherical harmonics to test spherical symmetry.  This
is analogous to the discrete Fourier expansion on a hypercubic lattice
to recover Poincar\'e invariance at long distances. However the
transform on a simplicial sphere is a bit more involved as explained
in detail in Appendix~\ref{app:Ylm} . The result is the expansion,
\be
\langle s(\hat n_i) s(\hat n_j) \rangle = \sum_{\ell', m'; \ell,m}
C_{\ell' m'; \ell ,m} Y^*_{\ell' m'}(\hat n_i) Y_{\ell m}(\hat n_j) \; ,
\ee
with $i,j = 1, \cdots ,N$. 
The coefficients are well approximate by 
  \begin{equation}
    C_{\ell' m'; \ell ,m} = \dfrac{1}{N^2}  \sum_{ij}  \sqrt{g_i g_j}
    Y_{\ell' m'}(\hat n_i) Y^*_{\ell, m}(\hat n_j) \< s_i s_j \> +
    O(a^2)\; ,
  \end{equation}
  for $\ell, \ell' \ll \sqrt{N}$. 
  In the continuum limit spherical symmetry demands that
  only the diagonal terms survive, 
  \be
  C_{\ell' m'; \ell ,m}    =  \delta_{\ell',\ell} \delta_{m',m} C_{\ell,  m} 
  \ee
  and with  $C_{\ell,  m}$  independent of m, the  scalar two point function only depends on the geodesic through
  $\hat n_i\cdot \hat n_j = \cos(\theta_{ij})$.  In practice we have only
  performed tests on  the m dependent diagonal term $\ell' = \ell, m' = m$
  for harmonics $\ell \le 12 $.

\begin{figure}[h]
    \centering
    \includegraphics[width=0.5\textwidth]{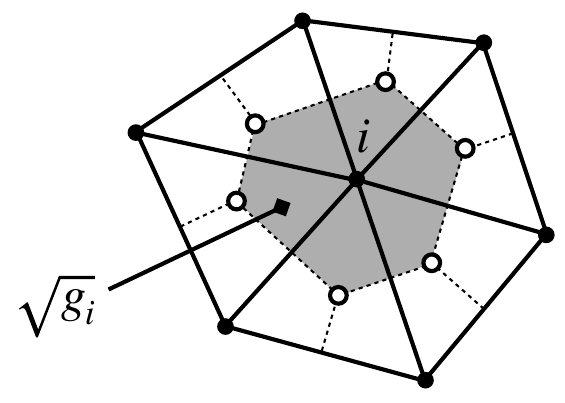}
    \caption{The shaded region is the Voronoi area, $\sqrt{g_i} = A_i$, at the lattice site $i$. The open circles are the circumcenters of the adjacent triangles.}
    \label{fig:voronoi}
\end{figure}

  Here, $N$ is the number of lattice sites and the quantity
  $\sqrt{g_i}$ is a discrete integration measure for each
  site. Following the circumcenter-based conventions of discrete
  exterior calculus~\cite{Desbrun2005DiscreteEC}, it is proportional to the area of the dual simplex associated with the site (i.e. the Voronoi area) as shown in Figure \ref{fig:voronoi}.
Because we are using a non-uniform lattices, we adopt a lattice
spacing definition as  a global average over all of the sites. For a lattice discretization of a sphere with $N = 2 + 10 L^2$ vertices, we simply define the lattice spacing to be equal to the square root of the average Voronoi area, 
\begin{equation}
\label{eq:a_lat_def}
\dfrac{a^2}{r^2} = \dfrac{1}{N} \sum_i \sqrt{g_i}
\end{equation}
where $r$ is the radius of the sphere.

\subsection{Restoring Rotational Symmetry}
\label{sec:RotationalSym}

The 2-sphere is invariant under the orthogonal group O(3), which has an infinite set of irreducible representations labeled by the familiar $\ell$ index used to describe the spherical harmonics, $Y_{\ell m}(\theta,\phi)$. For the spin-spin correlation function in our lattice model, restoration of rotational symmetry in the continuum limit requires that for each $\ell$, all of the measured $C_{\ell m}$ coefficients for $m \in \{-\ell, ..., \ell\}$ must become degenerate as $a \to 0$. To check this, we define a measurement of rotational symmetry breaking
\begin{equation}
\label{eq:symm_breaking_meas}
    \delta C_{\ell} = 1 - \dfrac{C_{\ell m}^{\textrm{(min)}}}{C_{\ell m'}^{\textrm{(max)}}}
\end{equation}
which captures the maximum deviation between the $C_{\ell m}$ coefficients for a given value of $\ell$.

Because our lattices have been constructed to be symmetric under a discrete subgroup of O(3), some of these coefficients will automatically be degenerate. The full octahedral symmetry group $O_h$ contains a 3-dimensional irreducible representation $T_{1u}$ which maps exactly onto the $\ell=1$ irreducible representation of O(3). Similarly, the full icosahedral symmetry group $I_h$ contains the 3-dimensional and 5-dimensional irreducible representation $T_{1u}$ and $H_g$ which map exactly onto the $\ell=1$ and $\ell=2$ irreducible representation of O(3), respectively. The corresponding coefficients of the 2-point function on lattices with octahedral or icosahedral symmetry do not break rotational symmetry (up to statistical errors), and therefore they provide a good indication for what the symmetry-breaking measurement should look like for an unbroken irreducible representation.

\begin{figure}[h]
    \centering
    \includegraphics[width = 0.49\textwidth]{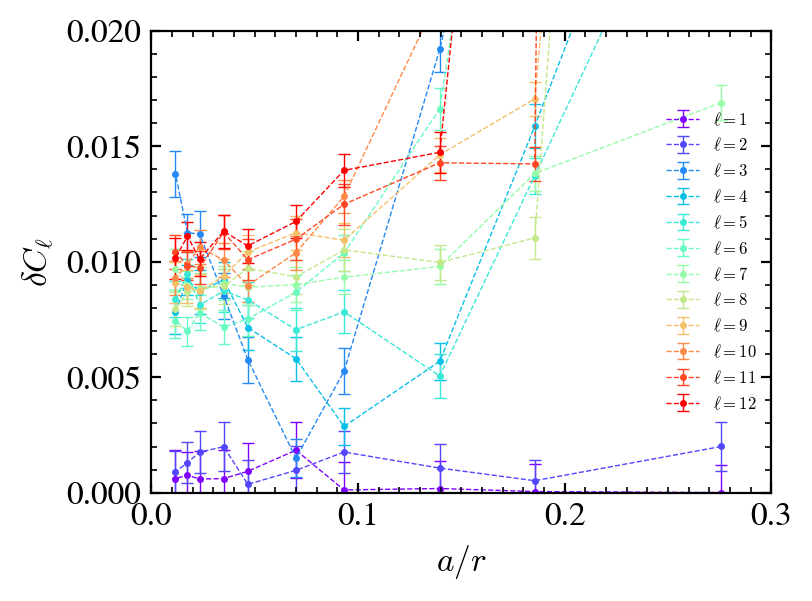}
     \includegraphics[width = 0.47\textwidth]{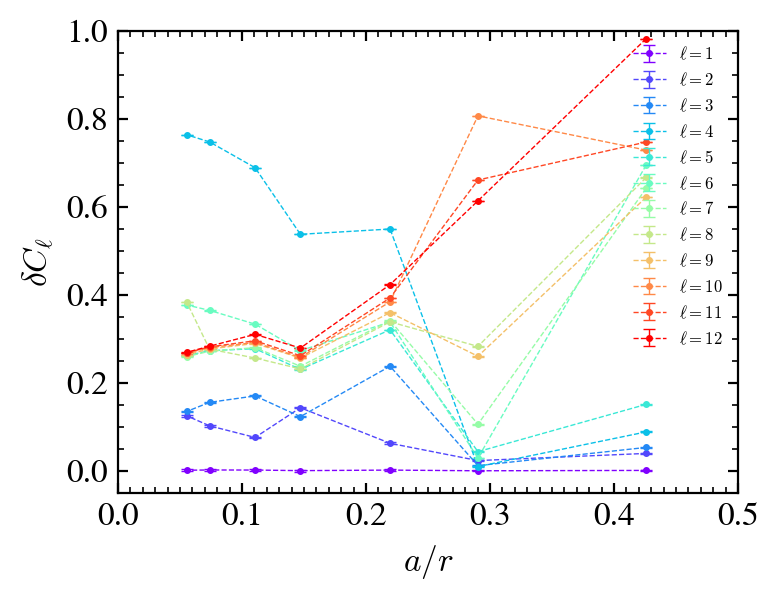}
    \caption{On left rotational symmetry breaking of the spin-spin
      correlation function using the basic icosahedral discretization
      of $S^2$ up to a refinement of 96. On the right severe breaking of rotational symmetry for Tegmark's octahedral based equal area pixelization of the celestial sphere \cite{Tegmark1996AnIM}.}
    \label{fig:q5_rot_break_naive}
  \end{figure}

In Fig. \ref{fig:q5_rot_break_naive} we show the rotational symmetry
breaking measurement as a function of the lattice spacing using the
basic discretization of the icosahedron. The measurement is clearly
approaching a nonzero value in the continuum limit for $\ell \geq 2$,
which indicates that this construction does not restore rotational
symmetry.

  Because the triangular faces of the basic discretization of the
  2-sphere fail to satisfy the constraints of equal circumradius and
  equal perimeter, it is perhaps unsurprising that the higher $\ell$
  coefficients of the spin-spin correlation function do not converge
  to zero in the continuum limit. One might have hoped that because
  the triangular faces in the basic discretization have \emph{locally}
  uniform circumradius and perimeter (i.e. deviations in the
  circumradius and perimeter of neighboring triangles go to zero in
  the continuum limit), this might have been sufficient to satisfy the
  geometric constraints required by our derivation in
  Sec.~\ref{sec:Uniformity}. However, it is clear from our
  measurements here that variations in triangle geometry over long
  distances result in a lattice theory which fails to fully restore
  continuum symmetries. We therefore conclude that the critical
  couplings derived in Sec.~\ref{sec:WMcontinuum} do not result in a
  lattice theory with a well-defined continuum limit if these
  geometrical constraints are only satisfied locally.

\begin{figure}[H]
    \centering
    \includegraphics[width=0.49\textwidth]{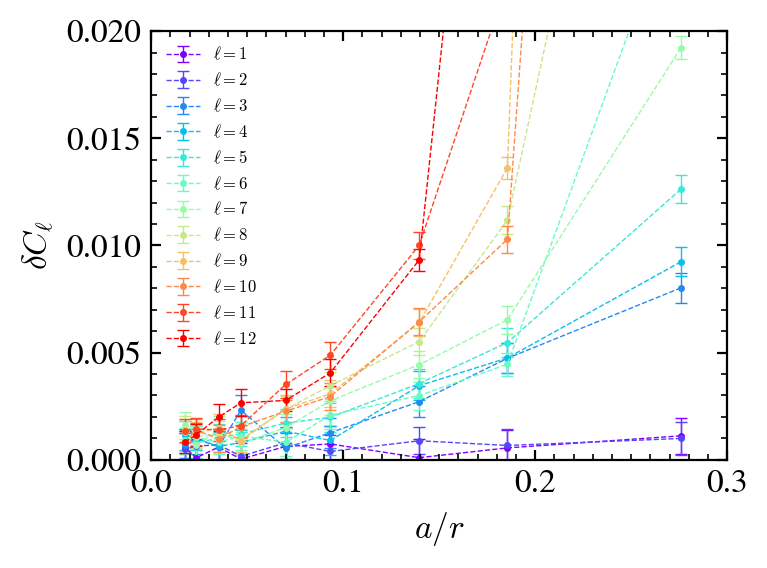}
     \includegraphics[width=0.49\textwidth]{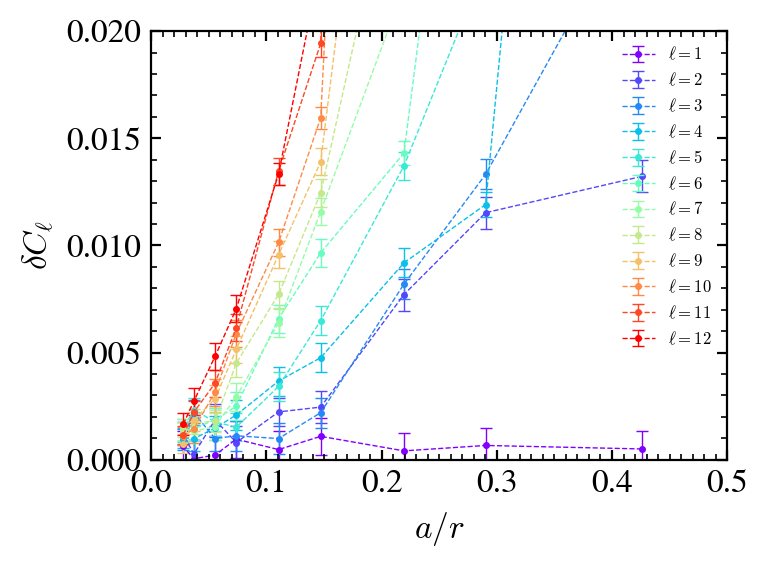}
    \caption{Rotational symmetry breaking of the spin-spin two-point function as a function of lattice spacing using the modified discretization up to a refinement of 64 for the sphere using an  icosahedral (left) compared to octahedral (right) base lattice .}
    \label{fig:rot_break_mod}
\end{figure}

\begin{figure}[H]
    \centering
    \includegraphics[width=0.42\textwidth]{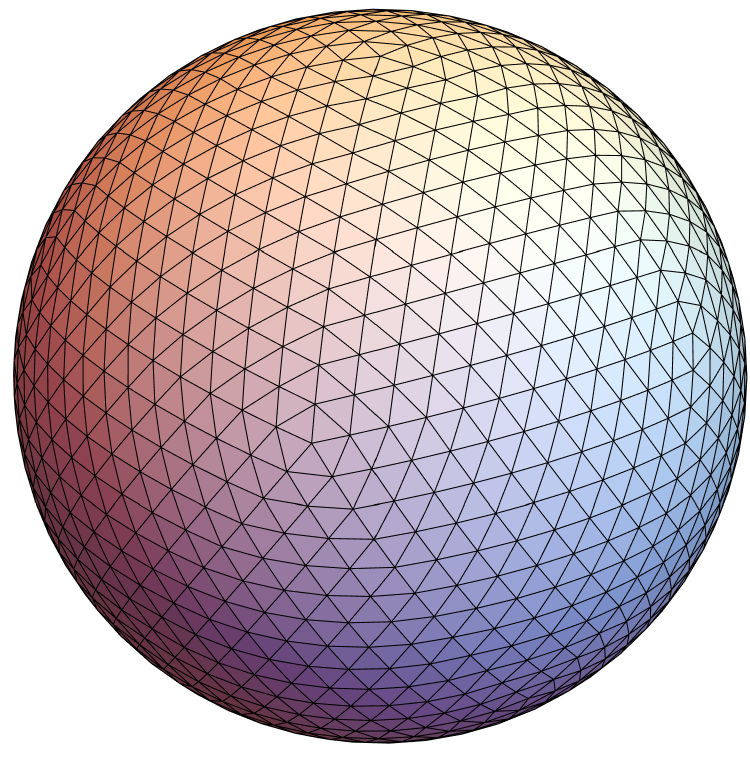}
     \hfill
    \includegraphics[width=0.42\textwidth]{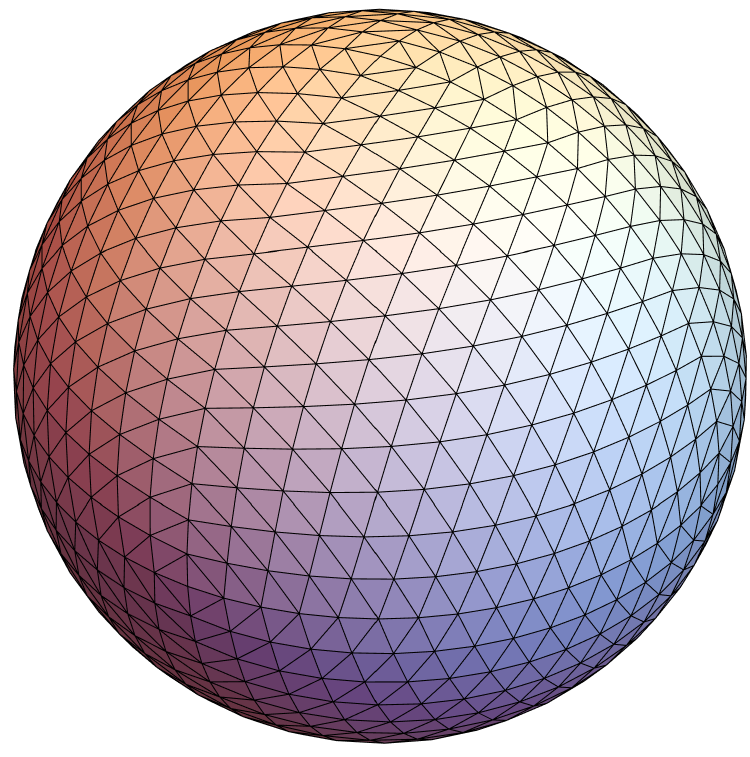}

    \caption{Simplicial discretizations of the sphere after applying the optimization procedure described in this appendix. On the left is the icosahedral lattice with a refinement of L= 12, compared to on the right the octahedral  lattice with a refinement of L =16.}
    \label{fig:lattice_uniform}
\end{figure}

To remedy this issue, we repeat the Monte Carlo simulations of the
previous section, but this time we use the modified lattice
construction described in Appendix \ref{appendix:iterative_uniform},
which explicitly minimizes non-uniformities in the circumradius and
perimeter of triangular faces so that \emph{global} variations in
these quantities go to zero in the continuum limit. We again plot the
symmetry breaking measurement $\delta C_{\ell}$ as a function of
lattice spacing in Fig. \ref{fig:rot_break_mod}.  For both the
octahedral and icosahedral discretizations, the rotational
symmetry-breaking measurement goes to zero for all measured
irreducible representation within statistical uncertainty. This result
indicates that global uniformity in the triangle geometry is necessary
to generate a sequence of lattices for which this model restores
rotational symmetry in the continuum limit. It is especially promising
that this procedure works even for the octahedral lattice, which
requires much larger variations in triangle shape in order to fully
tesselate the sphere, which can be seen explicitly in
Fig. \ref{fig:lattice_uniform} comparing the octahedral and
icosahedral examples.

\begin{figure}[H]
    \centering
    \includegraphics[width =  0.6\textwidth]{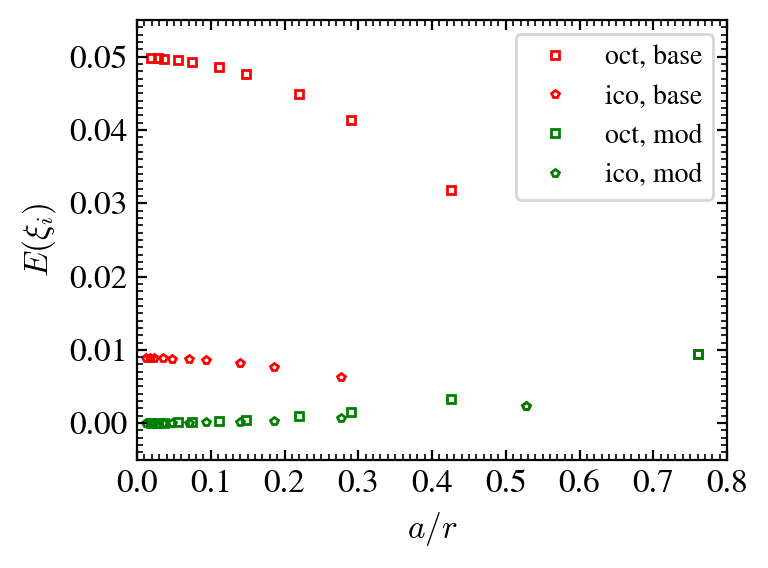}
    \caption{Non-uniformity measure for the octahedral and icosahedral discretizations of $S^2$ as a function of lattice spacing before and after being modified using the optimization procedure described in this appendix.}
    \label{fig:uniform}
\end{figure}

 For a quantitative comparison of the unmodified basic lattice and the modified lattice, in Fig. \ref{fig:uniform} we show the non-uniformity measure $E$ as a function of the lattice spacing as defined in Eq.~(\ref{eq:a_lat_def}) for both the octahedral and icosahedral lattices. We can clearly see that the non-uniformity in the unmodified lattice approaches a nonzero value in the continuum limit, whereas in the modified lattice the non-uniformity goes to zero roughly quadratically in the lattice spacing.

It is worth noting that although the triangle circumradius and
perimeter are the quantities that appeared in our derivation of the
Ising critical couplings, in general there are many other ways to
define the measure of uniformity for a simplicial lattice. As an
example, we tested our simulation of the Ising model on a discretized
sphere using the optimization procedure described in
\cite{Tegmark1996AnIM}, which adjusts the positions of the lattice
sites so that the Voronoi areas of all sites become approximately
equal. However, just as we saw when using the basic discretization,
this construction fails to restore rotational symmetry in the
continuum limit. The rotational symmetry breaking is especially strong
using a octahedral base lattice, as shown earlier in
Fig.~\ref{fig:q5_rot_break_naive}. We also tested several other
methods for adjusting the vertices with similar
results~\cite{Xu2006DiscreteLO,Ahrens2009RotationallyIQ,Iga2014ImprovedSA,Fornberg2014OnSH}. We
therefore conclude that the definition of uniformity based on the
circumradius and perimeter is indeed necessary to ensure a valid
continuum limit for the critical Ising model.

\begin{figure}[h]
   \centering
    \includegraphics[width =0.5\textwidth]{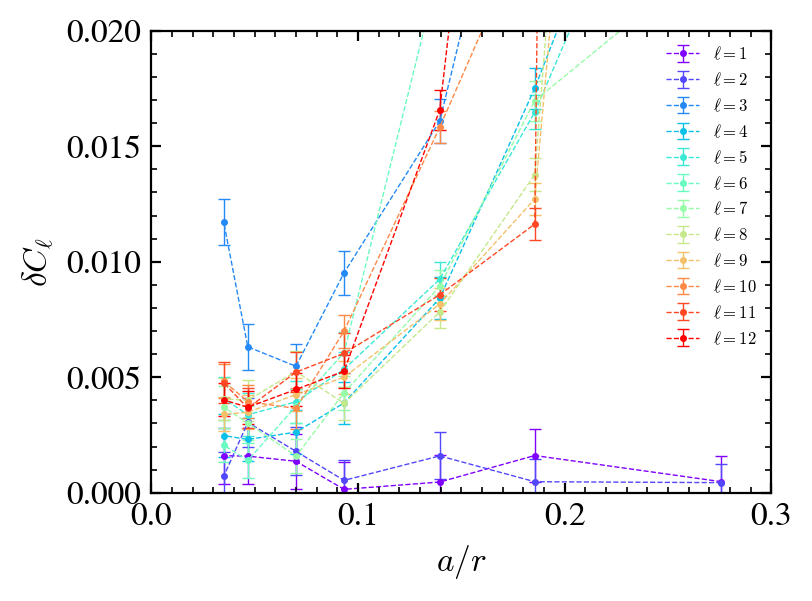}
    \caption{Breaking of rotational symmetry in critical $\phi^4$ theory with local perturbative counter-terms using the basic icosahedral discretization of $S^2$.}
    \label{fig:phi4_break}
  \end{figure}
  
  It is worth noting that the restoration of rotational symmetries in
  a lattice model of a sphere is a challenging problem which has been
  encountered in previous works. In \cite{Brower2012LatticeRQ}
  attempts were made to simulate the 3d Ising CFT on the manifold
  $\mathbb{R} \times \mathbb S^2$ of radial quantization using an
  icosahedral discretization of the sphere and uniform
  couplings. Although this construction was shown to have a
  well-defined critical point, rotational symmetry was not restored in
  the continuum limit. In
  \cite{Brower2018LatticeF,Brower2020RadialLQ}, simulations of
  critical $\phi^4$ theory were performed using a generalization of
  the finite element method. Again, it was shown that a well-defined
  critical point exists after adding a perturbative mass counter-term
  to account for the non-uniform UV divergence inherent in the
  lattice. However, as shown in Fig.~\ref{fig:phi4_break}, higher
  precision measurements of the scalar 2-point function reveal a
  slight breaking of rotational symmetry in the continuum limit, very
  similar to the same results presented in
  Fig.~\ref{fig:q5_rot_break_naive} for the Ising model on the basic
  discretization of the sphere. It therefore seems likely that some
  additional geometric constraints may be necessary in order to
  restore rotational symmetry in these theories as well. It's possible
  that the same constraints used here (uniform triangle circumradius
  and perimeter) may also work for other theories, but although we
  have begun to study this possibility it is unclear if this is the
  case.

\subsection{Agreement with the Ising CFT}
\label{sec:CFTtest}

\begin{figure}[H]
    \centering
    \includegraphics[width=0.49\textwidth]{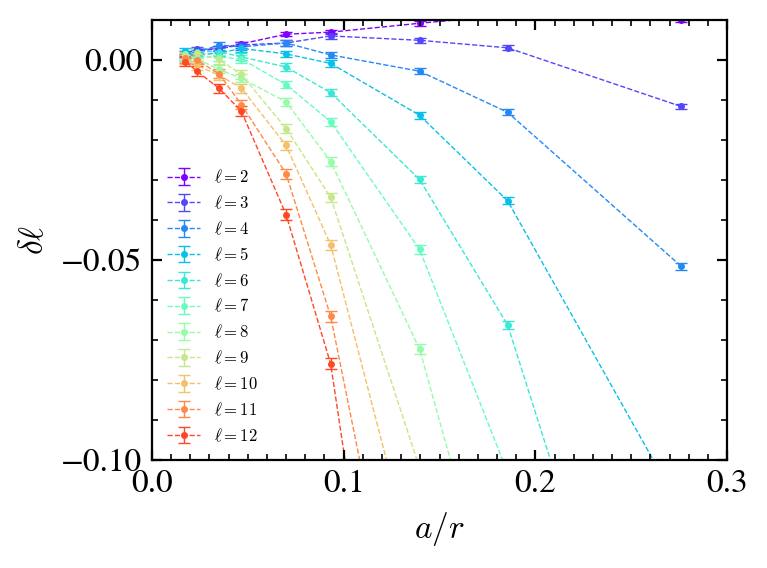}
    \includegraphics[width=0.49\textwidth]{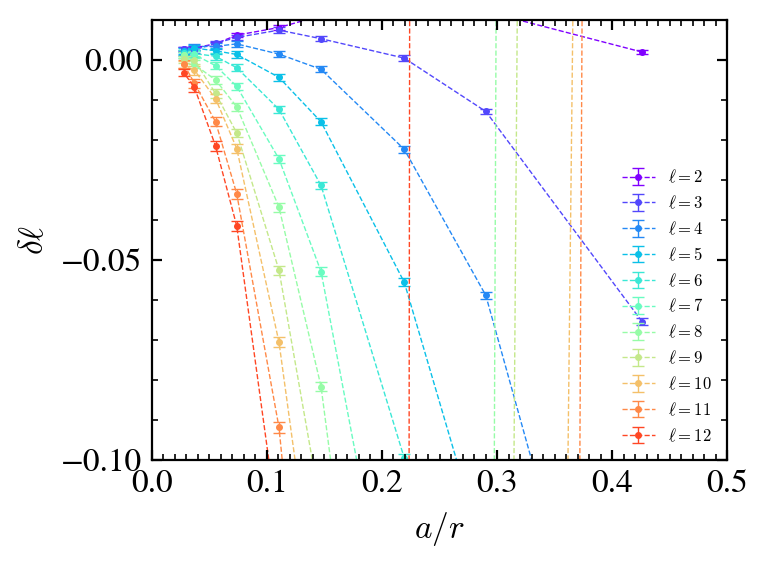}
    \caption{Conformal symmetry breaking measurement for the spin-spin two-point function as a function of lattice spacing using the modified discretization of the sphere comparing our  icosahedral (left) lattice to the octahedral (right)  lattice up to a refinement of $L = 64$.}
    \label{fig:cft_break_mod}
\end{figure}

Now that we have confirmed that the spin-spin correlation function in our lattice theory becomes rotationally symmetric in the continuum limit, we would like to check that it also agrees with the exactly known analytical result for the 2d Ising CFT on a 2-sphere: 
\begin{equation}
    \langle s_i s_j \rangle \propto \dfrac{1}{(1 - \hat n_i \cdot \hat n_j)^{\Delta_{s}}} = \sum_{\ell} \dfrac{2 \ell + 1}{2} F_{\ell} P_{\ell} (\hat n_i \cdot \hat n_j)
\end{equation}
where $\Delta_{s} = 1/8$ and we have expanded in a series of Legendre polynomials which have coefficients
\begin{equation}
\label{eq:2pt_coeff}
    \dfrac{F_{\ell}}{F_0} = \dfrac{\Gamma(\Delta_{s} + \ell) \Gamma(2 - \Delta_{s})}{\Gamma(\Delta_{s}) \Gamma(2 - \Delta_{s} + \ell)}
\end{equation}
which we can measure directly from lattice configurations generated by our Monte Carlo simulations. Eliminating $\Delta_{s}$ and solving for $\ell$ we find
\begin{equation}
    \ell = \dfrac{(F_0 - F_1)(F_{\ell-1} + F_{\ell})}{(F_0 + F_1)(F_{\ell-1} - F_{\ell})}\;.
\end{equation}
To check that the lattice simulation agrees with the analytic result, we first calculate a conformal symmetry breaking measurement
\begin{equation}
    \delta \ell = 1 - \dfrac{(F_0 - F_1)(F_{\ell-1} + F_{\ell})}{\ell (F_0 + F_1)(F_{\ell-1} - F_{\ell})}
\end{equation}
which should go to zero in the continuum limit. This quantity is plotted as a function of lattice spacing in Fig.~\ref{fig:cft_break_mod}, showing the expected behavior for both the octahedral and icosahedral lattice for all irrupts of O(3) up to $\ell = 12$.

\begin{figure}[h]
    \centering
    \includegraphics[width=0.49\textwidth]{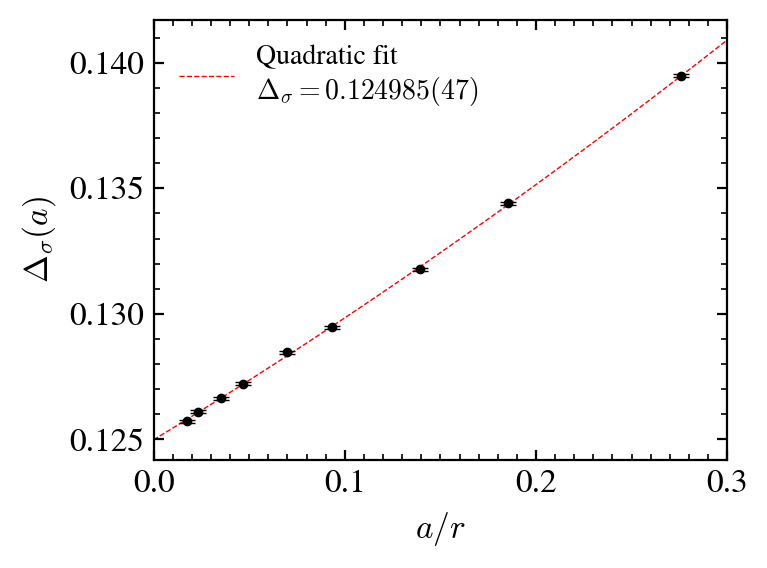}
    \includegraphics[width=0.49\textwidth]{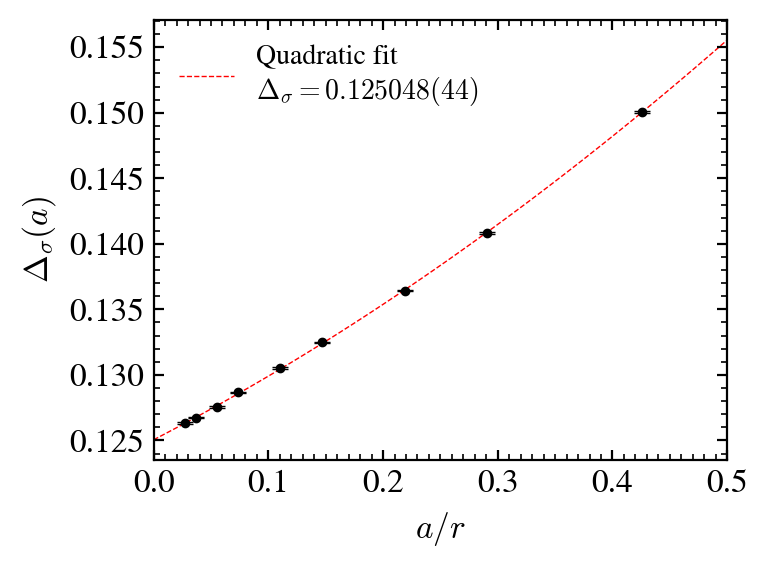}
    \caption{Continuum extrapolation of the scaling exponent $\Delta_{s}$ for the modified icosahedral (left) lattice compared to modified octahedral (right). }
    \label{fig:sphere_delta_sigma}
\end{figure}

Finally, we measure the scaling exponent of the $s$ operator on the lattice via
\begin{equation}
    \Delta_{s} = \dfrac{2 F_1}{F_1 + F_0} 
\end{equation}
which is plotted as a function of lattice spacing in Fig. \ref{fig:sphere_delta_sigma}. Extrapolating to the continuum limit by fitting to a quadratic polynomial in $a$, we obtain the results $\Delta_{s} = 0.125048(44)$ with $\chi^2/\text{dof}=1.8$ for the octahedral lattice and $\Delta_{s} = 0.124985(47)$ with $\chi^2/\text{dof}=1.8$ for the icosahedral lattice, both in excellent agreement with the exact value of $1/8$ and with a relative uncertainty of about $0.03\%$.

Our results indicate that by using the modified discretization of $S^2$, our lattice simulation accurately captures the properties of the lowest $\mathbb{Z}_2$-odd operator $s$. The 2d Ising CFT of course contains an infinite set of operators related by the Virasoro algebra. Measurement of additional operators using our lattice model is a subject for future study outside the present scope, however based on the high level of accuracy of our results and the delicate nature of a properly-tuned conformal field theory, we believe it is likely that our lattice formulation is sufficient to capture the properties of all of the operators in the 2d Ising CFT in the continuum limit, given sufficient computational resources.

Another approach taken recently is to use the so-called ``fuzzy sphere'' which projects the states of the theory in the basis of spherical harmonics instead of using a traditional spatial lattice~\cite{Zhu2022UncoveringCS,Hu2023OperatorPE}. In this case, the lattice cutoff is replaced by a maximum spherical harmonic quantum number, which is a more manifest way to ensure that rotational symmetry is restored in the continuum limit. This method shows promising results for the lowest operators of the 3d Ising CFT, though it is unclear how difficult it would be to generalize to other theories.

\label{sec:Conclusion}
\section{Conclusion}

We have demonstrated that there is a nearest neighbor Ising model
properly implemented on triangulation of the sphere which we
conjecture yields the  exact $c =1/2$ minimal Ising CFT in the
continuum limit.  This is based on both theoretical support and Monte
Carlo simulations. There are more stringent numerical tests and
theoretical arguments being pursued to support this conjecture.  We
are increasing the statistical accuracy for the $\<s_xs_y \>$
correlation function as well as computing the scalar 4 point function
and the correlator for the energy momentum tensor.  The energy
momentum tensor plays a crucial theoretical role as the exact marginal
operator conjugate to the metric tensor $g_{\mu \nu}$.  Our working
hypothesis is that smoothing the geometry is the first step to extract  the spherical differential from  a Regge piecewise flat manifold.

The benefit of the 2d Ising soluble theory is both its theoretical
foundation in our affine analysis in flat space in
Ref.\cite{Brower_2023} and comparison with the exact solution on
$\mS^2$.  The 3 state Potts model and SUSY tricritical Ising model
(TIM) provide interesting further examples to try.  To understand
this in a larger context beyond solvable lattice models, we are
developing similar analysis for $\lambda_0 \phi^4$ theory where the
map from the geometry of edge lengths $\ell_{ij}$ to coupling is no
longer known analytically.  We expect a smooth interpolation of the
map from FEM map~(\ref{eq:freeMap}) at zero bare coupling $\lambda_o$
to the Ising map (\ref{eq:uniform_couplings}) at infinite
coupling. But the general dependence on the bare coupling, $\lambda_0$, must be found numerically.  This is a first step in developing
efficient numerical algorithms of the affine map for general
renormalizable lattice field theories.  To go to higher dimensions, we
are starting with the 3d Ising theory in radial quantized form on
$\mR \times \mS^2$ and on the Riemann sphere, $\mS^3$, as simplicial
refinement of the 600 cell.

In conclusion it is useful to summarize our intuitive picture for the
central role of the affine parameter in constructing more general
solutions to lattice field theories on smooth curved manifolds. In
d-dimension a global affine transformation,
$ x^{\mu} = A_{\mu i} \xi^i+ b^\mu$, extends the $d(d+1/2)/2$
parameters of the Euclidean Poincar\'e group by an additional
$d(d+1)/2$  parameters. These are equivalent to a general metric
tensor, $ds^2 = g_{ij} d\xi^i d\xi^j = [A^T A]_{ij} d\xi^i d\xi^j$ on each
tangent plane.  The  discrete simplicial geometry has been extensively
studied~\cite{JWBarrett_1987} in the context of Regge Calculus on higher dimensional
triangular  lattices. Each piecewise flat simplex is
an affine subspace with the parameters represented by $d(d+1)/2$ edge
lengths of the d-simplex. The gauge potential (or Christoffel symbol)
define the parallel transport between adjacent simplicies across the
share boundary.

 In the quantum field theory the affine transformation
is generated by the $d(d+1)/2$ components of the conserved energy
momentum operator.  In a  conformal theory, it  is a  marginal
traceless operator. The trace providing the relevant operator, breaking
conformal invariance.  We believe the fundamental geometric problem
is constructing the map to the tangent plane,  re-introduce the differential
metric in the continuum limit of the Regge
manifold\cite{Feinberg:1984he}  that matches
the coordinate induced by the energy momentum
operator in the quantum field theory.  While extensions to
general lattice field theories are both non-trivial theoretically and
algorithmically challenging, we are optimistic that systematic sequence of
solutions for higher dimensional examples, including gauge and fermion
fields, will  systematically developed the requisite tools
with the ultimate goal of high performance 4d gauge theories leveraging methods for lattice QCD  at the Exascale.

\section*{Acknowledgements}
We thank   Cameron Cogburn, Liam Fitzpatrick, George Fleming,  Anna-Maria  Gl\"uck,  Ami Katz, Jin-Yun Lin,  Nobuyuki Matsumoto  and Chung-I Tan for very helpful discussions. This work was supported by the U.S. Department of Energy (DOE) under Award No.~DE-SC0019139 and Award No.~DE-SC0015845.

\bibliography{bib/IsingOnS2}

\appendix

\section{Simplicial  Spherical Harmonics}
\label{app:Ylm}

On a hypercubic lattice the analysis of translational and rotational symmetries using discrete
Fourier transform is obligatory. Here the analogous procedure
on a  triangulated sphere is the transform to discrete  spherical harmonics.
This requires a careful discussion of integration over a piecewise simplicial  manifold. 

 For example on a 2d simplicial manifold for the sphere,  the integral of a  smooth  function $\phi(x)$ with values
 $\phi_i$ on each vertex is defined
 by
 \be
\int d\Omega \phi(x) = \sum^N_{i =1} \sqrt{g_i} \phi_i + O(a^2) \ ,
\ee
where  the measure $ \sqrt{g_i}$ is the scaled circumcenter Delaunay
dual area: $\sum_i  \sqrt{g_i} = 4 \pi$. For any reasonable refinement
the error is $O(a^2)$ with with  $a \sim 1/\sqrt{N}$
given in Eq.~(\ref{eq:a_lat_def}). Indeed this is a
natural  generalization of the
trapezoidal rule on a  simplicial complex for $d > 1$. Many higher order
integration schemes are possible but are not needed for our current
analysis.

We now apply this to spherical harmonic, $Y_{lm}( \hat n_i)$,
evaluated on each site  on the sphere $\hat n_i = \vec r_i/R$.  
The latice addition theorem,
 \be
  P_{\ell}(\hat n_i \cdot \hat n_j) = \frac{4\pi}{2 \ell
    +1} \sum^{m = \ell}_{m = -\ell}Y^*_{\ell m}(\hat n_i) Y_{\ell ,
    m}(\hat n_j) \; ,
  \ee
  is still exact and the integration rule gives orthogonality, 
  \be
\sum_i \sqrt{g_i} Y^*_{\ell' m'}(\hat n_i) Y_{\ell m}(\hat n_i)  =  \delta_{\ell',\ell} \delta_{m',m} + O(a^2) \; ,
\ee
for  smooth modes with $\ell \ll L_{max} = O(1/a)$. 
This is applied to scalar two point function, 
\begin{equation}
  \< \widetilde \phi^*_{\ell' m'}  \widetilde  \phi_{\ell m} \> =  \sum_{\hat n_i} \sum_{\hat n_j} \sqrt{g_i}  
    Y^*_{\ell' m'}(\hat n_i) \langle \phi(\hat n_i) \phi(\hat
    n_j)\rangle  \sqrt{ g_j}  Y_{\ell m}(\hat n_j) \; .
  \end{equation}
  where   $\widetilde \phi_{lm} = \sum^N_{i = 1}\phi_i Y_{lm}(\hat
  n_i)$.
  The integration procedure allows us to invert  the expansion,
  \be
\langle \phi(\hat n_j)  \phi(\hat n_i) \rangle =   \sum_{\ell', m'; \ell,m}    C_{\ell' m'; \ell ,m} Y^*_{\ell' m'}(\hat n_i)  Y_{\ell m}(\hat n_j) \; ,
\ee
with expansion coefficient for low modes  modes $\ell, \ell' \ll L_{max}$ given by
  \begin{equation}
    C_{\ell' m'; \ell ,m} = \dfrac{1}{N^2}  \sum_{ij}  \sqrt{g_i g_j}  Y^*_{\ell' m'}(\hat n_i) Y_{\ell, m}(\hat n_j) \< \phi(\hat n_i) \phi(\hat n_j)\> +O (a^2) \; .
  \end{equation}
  Using the addition theorem,  spherical symmetry implies 
  \be
  C_{\ell' m'; \ell ,m}    =  \delta_{\ell',\ell} \delta_{m',m} C_{\ell} 
  \ee
  so that  in the continuum  $\langle \phi(\hat n_j)  \phi(\hat n_i) \rangle$
  is function the spherical geodesic through $\hat n_i \cdot \hat n_j = \cos(\theta_{ij})$.  In Sec.\ref{sec:projSphere}, 
  substituting $\phi(\hat n_i)
  \rightarrow  s_i $, this procedure applies equally well to  smooth quantum
  Ising  correlators  as one  approaches  a  second order critical surface.

 \section{Smoothing the Simplicial Manifold.}

 \subsection{Vertex degrees of freedom}

 On our graph illustrated in Fig.~\ref {fig:ico_refine} with
 refinement level $L$ has $F = 20 L^2$ faces, $E = 30 L^2$ edges
 and $N= 2 + 10L^2$ vertices. Embedded on the sphere the 3d unit vectors
 represent $2 N = 4 + 20 L^2$ dof.  After removing the 3 rotations and
 fixing the scale by the sphere's radius, the independent degrees of freedom are
 equal to the number of coupling constants $K_{ij}$ or the
 corresponding edge lengths $ \ell_{ij} = |r_i - r_j|$.

 In our implementation, we also seek to retain as much symmetry as possible, namely the discrete subgroup
 of $S(3)$ of the original platonic solid.  This can be accomplished by a judicious choice of the
 degrees of freedom of the problem, which we will describe here for an
 icosahedral discretization. It is important to note that the
 simplicial graph remains fixed during this procedure. This method
 works equally well for discretizations with octahedral or tetrahedral
 symmetry, and can also be generalized for discretizations of
 $S^3$. It is also possible to generalize to non-spherical manifolds,
 though we have not yet had a reason to do so.  

 We first identify ``orbits'' which are sets of vertices which
 transform into one another under the icosahedral group
 transformations. We use the full icosahedral group which includes
 rotations and reflections (120 group elements), but this method will
 also work for the chiral icosahedral group which includes only
 rotations (60 group elements). The number of distinct vertices in an
 orbit (the orbit's degeneracy) depends on whether its vertices lie on
 any symmetry axes of the icosahedron. For example, an orbit with
 vertices at the midpoints of the icosahedral edges has a degeneracy
 of 30 (one per icosahedral edge), while an orbit with vertices on one
 of the reflection axes of an icosahedral face has a degeneracy of 60
 (three per icosahedral face). An orbit which does not lie on any
 symmetry axes has a degeneracy of 120 (six per icosahedral face).

 We parameterize an orbit's position by its barycentric coordinates
 within an icosahedral face. The barycentric coordinates describe a
 position on an icosahedral face, which is then projected onto the
 unit sphere. Using barycentric coordinates allows us to define the
 coordinates of all vertices in an orbit simultaneously on all 20
 icosahedral faces. In addition, permuting the order of the 3
 barycentric coordinate values generates all of the vertices within a
 single icosahedral face.

 In order to preserve icosahedral symmetry, we require that orbits
 with vertices on a symmetry axis remains on that symmetry axis during
 the minimization procedure. Thus, though a point on a sphere has two
 degrees of freedom in general, this constraint reduces the number of
 degrees of freedom for some orbits (e.g. an orbit with vertices at
 the midpoints of the icosahedral edges has no degrees of
 freedom). The reduced set of orbit degrees of freedom, denoted
 $\xi_i$ where $i$ runs over all of the remaining degrees of freedom,
 can be freely adjusted without breaking icosahedral symmetry. Once we
 determine the values of $\xi_i$ which minimize $E$, we simply use the
 action of the icosahedral group elements to compute the vertex
 coordinates of all of the lattice sites in each orbit as described
 above.

 This reduction by imposing symmetries reduces  the dof continues to match
 the constraint of a single triangle term. However as described in the
 text we consider the over constrained system with both area (A) and
 perimeter (P) forms.  The use of over constraint optimization appears
 to be important for these highly non-linear optimization appears to
 be important.

\subsection{Iterative method for generating uniform spherical meshes}
\label{appendix:iterative_uniform}

In Sec. \ref{sec:RotationalSym}, we show that rotational symmetry of the
spin-spin correlation function was broken in the continuum limit for
simulations of the critical Ising model on the basic discretization of
a sphere. The failure to restore rotational symmetry in the continuum
limit can be traced back to the fact that our derivation of the
critical couplings on a simplicial lattice required all of the
triangles to have equal circumradius and perimeter, which is not the
case for the basic discretization of the sphere. In this appendix I
will describe an iterative method for modifying the basic
discretization of $S^2$ to make the triangles more uniform.

After identifying the reduced set of degrees of freedom $\xi_i$, we
use Newton's method to solve the non-linear system of equations
\begin{equation}
    \dfrac{\partial E}{\partial \xi_i} = 0\;.
\end{equation}
We use the barycentric coordinates of the vertices from the basic discretization as the initial guess $\xi_i^{(0)}$, then the $(k+1)$-th iteration of Newton's method~\cite{Dennis1983NumericalMF} sets
\begin{equation}
    \xi_i^{(k+1)} = \xi_i^{(k)} - \left[ \dfrac{\partial^2 E^{(k)}}{\partial \xi_i^{(k)} \partial \xi_j^{(k)}} + \mu^{(k)} \delta_{ij} \right]^{-1} \dfrac{\partial E^{(k)}}{\partial \xi_j^{(k)}}
\end{equation}
where $E^{(k)}$ is the non-uniformity measure computed on the lattice with the vertex coordinates determined by the orbit coordinates $\xi_i^{(k)}$. We compute the partial derivatives via first-order finite differences with a step size of $10^{-5}$, which is efficient for double-precision floating point numbers. The parameter $\mu^{(k)}$ is a preconditioning factor and is chosen to ensure that $E^{(k)}$ is strictly decreasing for successive iterations. We continue the iterative process until the quantity $1 - E^{(k+1)} / E^{(k)}$ becomes less than $10^{-10}$.

\end{document}

%% file: Figures/Duality2_rcb.tex
\begin{tikzpicture}[scale=2.0]


\draw[thick] (-0.433,0) -- (1.732+0.433,0);
\draw[thick] (0.433,1.5) -- (2.598+0.433,1.5);
\draw[thick] (-0.433/2,-0.375) -- (0.866*1.25,1.5+0.375);
\draw[thick] (1.732+0.433/2,-0.375) -- (0.866*0.75,1.5+0.375);
\draw[thick] (1.732-0.433/2,-0.375) -- (2.598+0.433/2,1.5+0.375);
\draw[thick] (2.598-0.433/2,1.5+0.375) -- (2.598+0.433/2,1.5-0.375);
\draw[thick] (0.433/2,-0.375) -- (-0.433/2,0.375);

\fill (0,0) circle[radius=0.0833];
\fill (0.866,1.5) circle[radius=0.0833];
\fill (1.732,0) circle[radius=0.0833];
\fill (2.598,1.5) circle[radius=0.0833];

\node[anchor=south] at (0.866,0.0) {$K_1$};
\node[anchor=south] at (1.732,1.5) {$K_1$};
\node[anchor=south east] at (0.5,0.65) {$K_3$};
\node[anchor=north west] at (2.1,0.9) {$K_3$};
\node[anchor=south west] at (1.25,0.65) {$K_2$};

\end{tikzpicture}

%% file: Figures/SimplicialCouplings.tex
\begin{tikzpicture}[scale=4.5]

\coordinate (v1) at (0,0);
\coordinate (v2) at (0,1);
\coordinate (v3) at (-0.7, 0.4);
\coordinate (v4) at (0.8, 0.6);

\coordinate (e12) at (0, 0.5);
\coordinate (e13) at (-0.35, 0.2);
\coordinate (e14) at (0.4, 0.3);
\coordinate (e23) at (-0.35, 0.7);
\coordinate (e24) at (0.4, 0.8);

\coordinate (c1) at (-0.178571, 0.5);
\coordinate (c2) at (0.25, 0.5);

\coordinate (n13) at (-0.0545367, 0.717061);
\coordinate (n23) at (-0.0158736, 0.310186);
\coordinate (n123) at (-0.428571, 0.5);
\coordinate (n14) at (0.1, 0.7);
\coordinate (n24) at (0.138197, 0.276393);
\coordinate (n124) at (0.5,0.5);

\draw[thick] (v1) -- (v2);
\draw[thick] (v1) -- (v3);
\draw[thick] (v1) -- (v4);
\draw[thick] (v2) -- (v3);
\draw[thick] (v2) -- (v4);

\draw[thick, densely dashed] (c1) -- (c2);
\draw[thick, densely dashed] (c1) -- (e13);
\draw[thick, densely dashed] (c1) -- (e23);
\draw[thick, densely dashed] (c2) -- (e14);
\draw[thick, densely dashed] (c2) -- (e24);


\draw [thick,fill=black] (v1) circle[radius=0.025];
\draw [thick,fill=black] (v2) circle[radius=0.025];
\draw [thick,fill=black] (v3) circle[radius=0.025];
\draw [thick,fill=black] (v4) circle[radius=0.025];

\draw [thick,fill=white] (c1) circle[radius=0.025];
\draw [thick,fill=white] (c2) circle[radius=0.025];

\node at (0.0, -0.1) {$i$};
\node at (0.0, 1.1) {$j$};
\node at (-0.25, 0.5) {$i^*$};
\node at (0.35, 0.5) {$j^*$};

\node[fill=white] at (0,0.7) {$K_{ij}$};
\node[fill=white] at (0.12,0.5) {$L_{ij}$};



\end{tikzpicture}

%% file: Figures/GenericIsing.tex
\begin{tikzpicture}[scale=4.5]

\coordinate (v1) at (0,0);
\coordinate (v2) at (0,1);
\coordinate (v3) at (-0.7, 0.4);
\coordinate (v4) at (0.8, 0.6);

\coordinate (e12) at (0, 0.5);
\coordinate (e13) at (-0.35, 0.2);
\coordinate (e14) at (0.4, 0.3);
\coordinate (e23) at (-0.35, 0.7);
\coordinate (e24) at (0.4, 0.8);

\coordinate (c1) at (-0.178571, 0.5);
\coordinate (c2) at (0.25, 0.5);

\coordinate (n13) at (-0.0545367, 0.717061);
\coordinate (n23) at (-0.0158736, 0.310186);
\coordinate (n123) at (-0.428571, 0.5);
\coordinate (n14) at (0.1, 0.7);
\coordinate (n24) at (0.138197, 0.276393);
\coordinate (n124) at (0.5,0.5);

\draw[thick] (v1) -- (v2);
\draw[thick] (v1) -- (v3);
\draw[thick] (v1) -- (v4);
\draw[thick] (v2) -- (v3);
\draw[thick] (v2) -- (v4);

\draw[thick, densely dashed] (c1) -- (c2);
\draw[thick, densely dashed] (c1) -- (e13);
\draw[thick, densely dashed] (c1) -- (e23);
\draw[thick, densely dashed] (c2) -- (e14);
\draw[thick, densely dashed] (c2) -- (e24);

\draw[densely dotted] (c1) -- (n13);
\draw[densely dotted] (c1) -- (n23);
\draw[densely dotted] (c1) -- (n123);
\draw[densely dotted] (c2) -- (n14);
\draw[densely dotted] (c2) -- (n24);
\draw[densely dotted] (c2) -- (n124);

\draw [thick,fill=black] (v1) circle[radius=0.025];
\draw [thick,fill=black] (v2) circle[radius=0.025];
\draw [thick,fill=black] (v3) circle[radius=0.025];
\draw [thick,fill=black] (v4) circle[radius=0.025];

\draw [thick,fill=white] (c1) circle[radius=0.025];
\draw [thick,fill=white] (c2) circle[radius=0.025];

\node at (-0.178571, 0.4) {$i$};
\node at (0.25, 0.4) {$j$};


\node at (-0.19, 0.65) {$\alpha_{1}$};
\node at (-0.32, 0.57) {$\alpha_{2}$};
\node at (-0.31, 0.43) {$\alpha_{3}$};

\node at (0.23, 0.67) {$\beta_{1}$};
\node at (0.40, 0.57) {$\beta_{2}$};
\node at (0.41, 0.43) {$\beta_{3}$};

\end{tikzpicture}